 \documentclass[final,5p,times,twocolumn,authoryear]{elsarticle}

\usepackage{geometry}
\usepackage{hyperref}
\hypersetup{
    colorlinks=true,
    linkcolor=blue,
    filecolor=magenta,      
    urlcolor=cyan,
}

\usepackage{xcolor}

\usepackage{amssymb}
\usepackage{booktabs}

\usepackage{footnote}
\usepackage{lineno}

\journal{Icarus}

\begin{document}

\begin{frontmatter}



\title{A photochemical model of ultraviolet atomic line emissions in 
the inner coma of comet 67P/Churyumov-Gerasimenko}


\author{Susarla Raghuram\corref{cor1}\fnref{label1}}
 \fntext[label2]{Corresponding author email address: 
 raghuramsusarla@gmail.com}
\author{Anil Bhardwaj}
\address{Physical Research Laboratory, Ahmedabad, 380009, India.}

\begin{abstract}
Alice  ultraviolet spectrometer onboard Rosetta 
space mission observed several spectroscopic emissions 
emanated from volatile species of comet 
67P/Churyumov-Gerasimenko (hear after 67P/C-G) during its 
entire escorting phase. The measured emission intensities,
 {when the comet was at around 3 AU pre-perihelion}, have been 
used to derive electron densities in the cometary coma 
assuming that  H~{\scriptsize{I}} and O~{\scriptsize{I}} 
lines are solely produced by  electron impact dissociative 
excitation of cometary parent species \citep{Feldman15}. We 
have developed a photochemical model for comet 67P/C-G to study the 
atomic hydrogen (H~{\scriptsize{I}} 1216, 
1025, \& 973 \AA), oxygen (O~{\scriptsize{I}} 1152, 
1304, \& 1356 \AA), and  carbon (C~{\scriptsize{I}} 1561 
\& 1657 \AA) line emissions by accounting for major  
production pathways.  The developed model has been used to 
calculate the emission intensities of 
these lines as a function of nucleocentric projected 
distance and also along the nadir view by varying the  
input parameters, viz., neutral abundances and cross 
sections. We have quantified the percentage contributions 
of photon and electron impact dissociative excitation 
processes to the total intensity of the emission lines, 
which has an important relevance for the analysis of 
 {Alice} observed spectra. It is found that in 
comet 67P/C-G, which is having neutral gas production rate 
of about 10$^{27}$ s$^{-1}$ when it was at 1.56 AU from 
the Sun, photodissociative excitation processes are  more 
significant compared to electron impact reactions in 
determining the atomic emission intensities.
Based on our model calculations, we suggest that the 
observed atomic hydrogen, oxygen, and carbon  emission 
intensities   can be used to  derive H$_2$O, O$_2$, 
and CO, abundances, respectively,  rather than electron 
density in the coma of 67P/C-G, when comet has a gas 
production rate  of $\ge$ 10$^{27}$ s$^{-1}$. 
\end{abstract}


\end{frontmatter}

\section{Introduction} 
The recent Rosetta  space mission, orbited, 
escorted, rendezvoused  
comet  {67P/Churyumov-Gerasimenko} (hereafter 
67P/C-G) and unravelled many unexplored details, with an
unprecedented spatial and temporal resolution, which were 
not possible by earlier missions. 
 {Alice} ultraviolet spectrometer onboard Rosetta is
designed to observe various spectroscopic emissions in the
wavelength region 700--2050~\AA\ and has studied the 
evolution of neutral environment around comet 67P/C-G  
during its entire  escorting phase, i.e. 2014 
August to 2016 September \citep{Stern07,Feldman15, 
Feldman16, Feldman18, Chaufray17,Noonan18}. The initial 
observations made at near-nucleus on comet 67P/C-G, during 
the end of September to November 2014 when the comet was at 
around 3 AU from the Sun, have shown many atomic spectral 
features \citep{Feldman15}. By observing H~{\scriptsize{I}} 
(Lyman-$\alpha$\ : 1216~\AA, Lyman-$\beta$\ : 1026~\AA, 
\& Lyman-$\gamma$\ : 972~\AA), C~{\scriptsize{I}} (1561 
\& 1657 \AA) and O~{\scriptsize{I}} (1152, 1304, \& 
1356~\AA) emissions on 67P/C-G, \cite{Feldman15} 
suggested that suprathermal electron impact dissociative 
excitation of neutral species  is the main generation 
process for these atomic line emissions. \cite{Chaufray17} have 
analysed  {Alice} spectra 
obtained during 2014 December, 2015 March, and 2015 May, 
when comet 67P/C-G was at {around} 3, 2, and 1.5 AU from the Sun, 
respectively, and concluded that suprathermal electron 
density is not varying as a function of cometocentric 
radial distance. Besides atomic H~{\scriptsize{I}}, 
O~{\scriptsize{I}} and C~{\scriptsize{I}} emissions, 
\cite{Feldman18} have recently reported fourth positive, 
Cameron band, and Hopefield-Birge band emissions of CO 
molecule   {from} the  2015 January 30 and 2016 May 14 observed 
spectra when the comet was at 2.5 AU pre-perihelion and 3 
AU  post-perihelion, respectively.

Rosetta Orbiter Spectrometer for Ion and 
Neutral Analysis (ROSINA) onboard Rosetta  was continuously 
monitoring the evolution of gaseous environment  during the 
entire mission period using both the spectrometers, viz., 
Double Focusing Mass Spectrometer (DFMS) and 
Reflection-type Time-Of-Flight (RTOF), as well 
as COmet Pressure Sensor (COPS). ROSINA could study the 
dynamical variation of the 
neutral and ion environment of cometary coma with the 
highest ever possible temporal and spatial resolution 
\citep{Balsiger07,Hassig15,Leroy15,Hoang17,	Fougere16, 
Bieler15, Gasc17b}. The first-ever in-situ detection of 
molecular oxygen in comets is one of the surprising results 
from the ROSINA observations. The ROSINA-DFMS neutral 
density measurements from September 2014 to April 2015, 
when the comet  was moving from 3.2 to 2 AU towards the 
Sun, have shown that  O$_2$ relative abundances are  
varying between 0.5 and 15\% with respect to H$_2$O 
\citep{Bieler15}.

By making 
in-situ measurements around 67P/C-G, the evolution of 
cometary ionospheric parameters was monitored 
by Rosetta Plasma Consortium (RPC) instruments. Ion and 
Electron 
Sensor (IES), which {is} a part of RPC, is designed to study the 
three-dimensional distribution of both ions and electrons 
\citep{Burch07}. The initial RPC-IES {measurements} of 
suprathermal electron spectra, when the comet was around 3 
AU, have shown that these electrons are accelerated to 
several hundred electron volts of energy in the cometary 
coma \citep{Clark15}. The 
plasma studies of  \cite{Galand16}  at 3 AU have also shown 
that 
suprathermal electrons play a  {more} significant role than solar 
photons in controlling the cometary ionosphere. But
 {the role of  solar photons increased when 
the comet was moving towards the Sun  and was} significant compared to 
suprathermal 
electrons 
in the cometary coma \citep{Heritier17, Heritier18}.

Since the observed various atomic emissions depend 
on  {neutral} species distribution in the cometary coma and also 
 {different energetic particles}, such as solar 
photons and suprathermal electrons, it is essential to 
model the emission intensities by accounting for all 
production and loss mechanisms of various excited species 
to understand the photochemistry of cometary coma. Several excitation 
processes have been proposed in the earlier modelling works for 
H~{\scriptsize{I}}, O~{\scriptsize{I}}, and 
C~{\scriptsize{I}} emissions \citep{Combi04, Combi98, 
Bhardwaj96, Bhardwaj99a}. Photon and electron impact 
dissociative excitation and solar 
resonance fluorescence  of cometary neutrals  {were} considered to be 
major 
excitation mechanisms for all the allowed transitions 
\citep{Feldman04}.

In this paper, we have modelled various Alice observed 
ultraviolet atomic
hydrogen (H~{\scriptsize{I}} Lyman emissions at 1216, 1026, 
\& 972 \AA), carbon (C~{\scriptsize{I}} 1561 \& 1657 
\AA), and oxygen (O~{\scriptsize{I}} 1152, 1304, \& 1356 
\AA) emissions by accounting for major excitation mechanisms, viz.,
photon and electron impact initiated reactions and solar resonance 
fluorescence of cometary species. 
The main aim of the current work is to 
quantify the 
contributions of electron and photon impact dissociative 
excitation processes in 67P/C-G coma, which produce the 
 {Alice} observed spectral lines.
The required model  input 
parameters, viz., the neutral composition 
of cometary coma, solar photon flux, photon and electron impact 
cross sections of major species are 
described in Section~\ref{sec:model-inputs}.
In Section \ref{model-des}, we have described the 
method of calculation for suprathermal thermal electron 
flux 
and the  ultraviolet emission intensities along 
the Alice line of sight. More details of calculation of 
steady-state suprathermal electron spectra in the 
cometary coma can be found in our 
earlier work \citep{Bhardwaj99a, Bhardwaj95, 
Bhardwaj96, Bhardwaj02, Bhardwaj11, Bhardwaj12}. 
The major results obtained from the model calculations
are presented in 
Section~\ref{sec:results}. In this 
section,  {we have also studied the sensitivity of the calculated 
intensities to several parameters : } neutral abundances and cross
 sections.  {These results are discussed in 
 Section~\ref{sec:discussion}}.  The 
complete work has been summarized and concluded 
in Section~\ref{summary}.   

\section{Model input parameters}
\label{sec:model-inputs}
\subsection{Neutral composition of cometary coma} 

In the model, H$_2$O, CO$_2$, CO, and O$_2$ have been
considered as primary neutral constituents of the  coma. Based on the 
ROSINA in-situ measurements, H$_2$O 
production rate is taken as 2 $\times$ 10$^{27}$ s$^{-1}$ 
and CO$_2$, CO  and O$_2$ relative abundances
with respect to water  are considered as 2.5, 2, 
and 4\%, respectively,  \citep{Hansen16, Hassig15, Bieler15}. 
This model input chemical composition resembles the   
cometary neutral environment of 67P/C-G, when it was at 1.56 AU from the Sun
on 25 May 2015, during {which Alice made spectroscopic observations 
\citep{Chaufray17}}.  
We have also calculated nadir intensities of different 
atomic emissions when the comet was at 1.99 AU. The variation
in the ROSINA-DFMS measured gas production rate and 
relative abundances of primary species are taken from 
\cite{Fougere16}, \cite{Hansen16}, and \cite{Hoang17}
and tabulated in Table~\ref{tab:neu}. 
Based on the ROSINA-DFMS measurements,
we have varied the 
relative abundances and gas production rates of the neutrals 
in our model to discuss the effect on the
calculated H~{\scriptsize{I}}, 
O~{\scriptsize{I}}, and C~{\scriptsize{I}}  
emission intensities.

\begin{table*}[htb]
	\caption{The volatile distribution around 67P/C-G
		at different heliocentric distances as observed by 
ROSINA.}	
	\centering
	\begin{tabular}{l c c c c c c c}
		\hline
		Date  & r$^a$ & H$_2$O$^b$  & 
		CO$_2$$^c$ & CO$^c$ &  
		O$_2$$^c$ \\
		&  (AU) & (s$^{-1}$) $\times$ 10$^{26}$  & (\%) & 
(\%) 	
		&  (\%) \\
		\hline
		2015 Mar. 29 & 1.99 & 1.8--12.5 & 1.6--4& 0.7--2 
		& 1.5   \\
		2015 May 15 & 1.56 & 15.3--80 & 3--6 & 
		1--7 & 5  \\
		\hline
	\end{tabular}
	
	\footnotesize{$^a$ Heliocentric distance;}
	\footnotesize{$^b$Water production rates are taken from 
{\cite{Hansen16}}; $^c$CO, CO$_2$ and O$_2$ are relative 
	abundances, with respect to water, are taken 
	from {\cite{Fougere16}};} 
	\label{tab:neu}
\end{table*}

\subsection{Solar radiation flux} 
Solar radiation flux was obtained from the Solar 
EUV Experiment (SEE) instrument onboard Thermosphere 
Ionosphere Mesosphere Energetics and Dynamics (TIMED) 
spacecraft measurement in the wavelength region 5--1900 \AA\ 
\citep[{http://lasp.colorado.edu/see/};][]{Woods05} 
and has been scaled as a function of an inverse square 
of heliocentric distance of the comet.  {We took into account 
the comet-Sun-Earth phase angle to extrapolate the 
solar flux measurements to the comet} for the given day of Alice observation.
\subsection{Photon and electron impact cross sections} 
\begin{figure}[htb]
	\centering
	\includegraphics[width=\columnwidth]{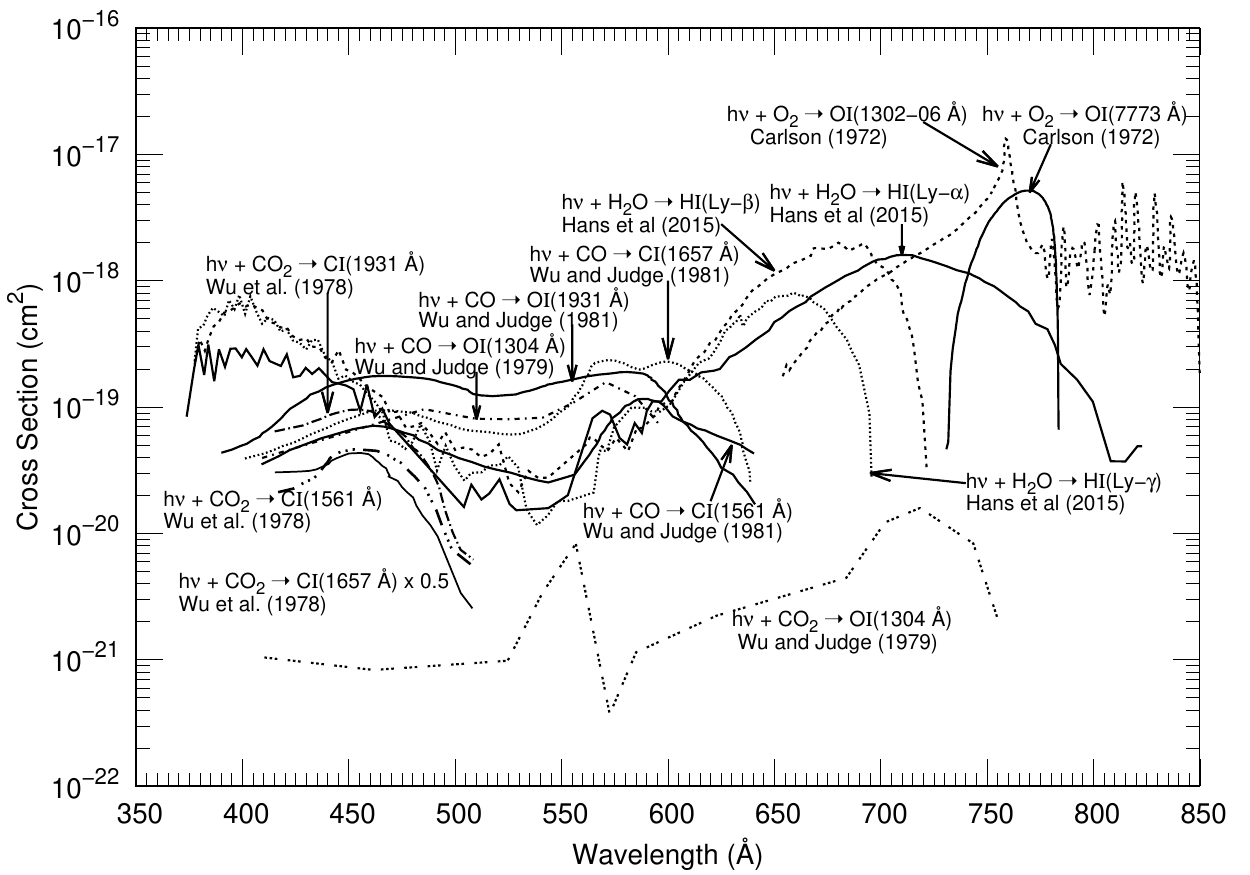}
	\includegraphics[width=\columnwidth]{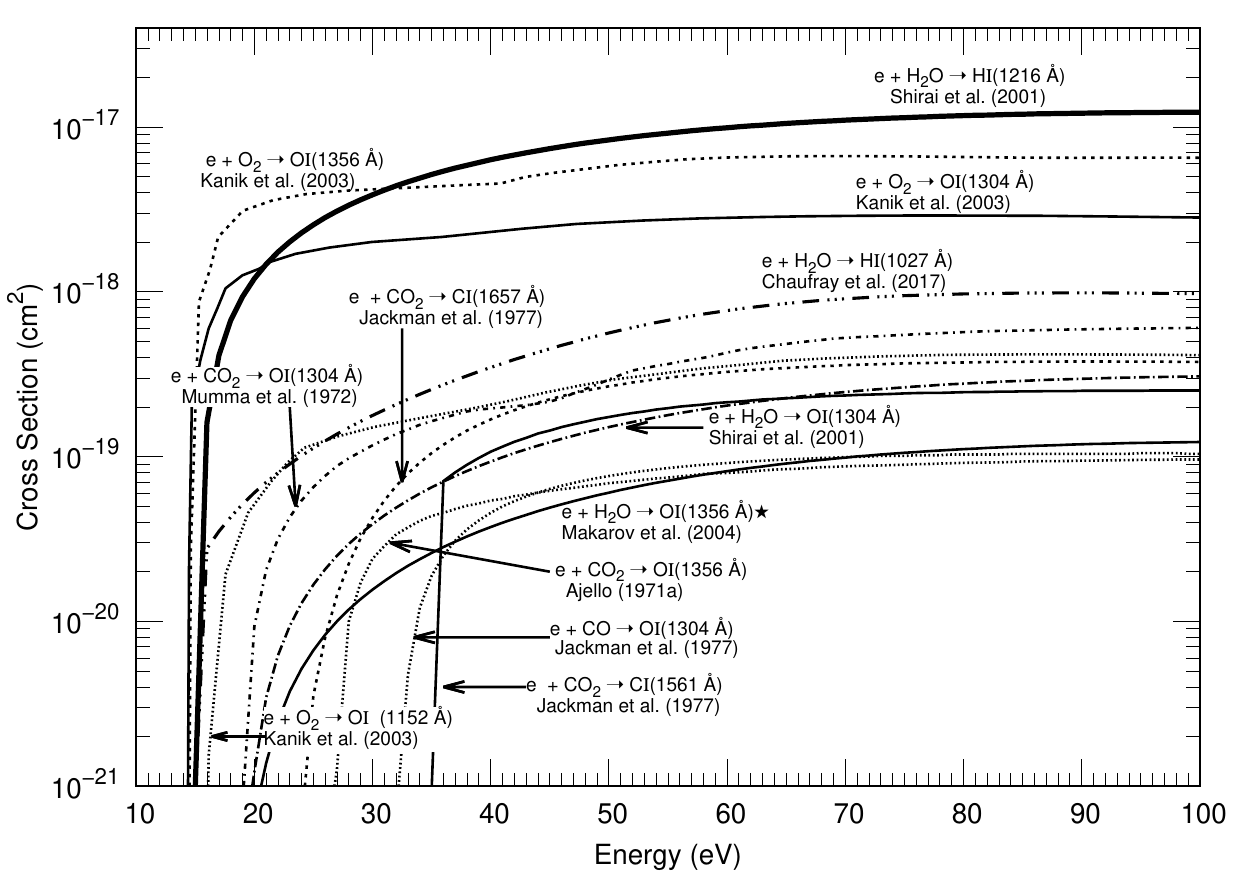}
	\caption{Model input  photon (upper panel) and 
		electron  (lower panel) impact dissociative 
		excitation cross sections  of neutrals
		producing atomic hydrogen, oxygen and carbon
		emissions.  Photodissociative excitation cross 
		section of CO$_2$\ producing  C~{\scriptsize{I}} 1657~\AA\ 
		emission
		is multiplied by a factor 0.5 to avoid 
		overlap of curves.
		$\bigstar$ Estimated cross 
		section based on \cite{Makarov04} measurement at 100 eV 
		by (see main text 
		for more details). h$\nu$ is photon and e 
		is electron.} 
	\label{fig:ele-csc}
\end{figure}

Total photoabsorption and photoionization cross 
sections for primary cometary species are taken from 
\cite{Huebner92}. 
The  electron impact total inelastic cross sections
of H$_2$O, CO$_2$\ and CO  are  taken from 
\cite{Shirai01, Itikawaco2, Liu94}, respectively. 
 {The photon and electron dissociative emission cross sections of neutral 
species are presented in Figure~\ref{fig:ele-csc}.}
Due to the lack of measured cross 
sections, several assumptions are made to incorporate some 
important dissociative excitation 
processes. The impact of our assumed cross sections on the 
calculated emission intensities will be discussed in  
 {Section~\ref{sec:effect-inputs}.}

\subsubsection{Cross sections for atomic hydrogen 
	Lyman emissions:  H~{\scriptsize{I}} 1216, 1026, 972 \AA}
 {Cross section for the emissions of H~{\scriptsize{I}} (Lyman-$\alpha$, 
Lyman-$\beta$, \& Lyman-$\gamma$) by 
photodissociation of H$_2$O between 375--825 \AA\ were taken from \cite{Hans15} 
and are in agreement with the earlier measurements of \cite{Wu88}.
}
By reviewing various measured electron impact cross sections
of H$_2$O, \cite{Itikawah2o} recommended \cite{Morgan74}
reported values for H~{\scriptsize{I}} Lyman-$\alpha$\ emission.
\cite{Shirai01} analytically fitted various  
atomic  hydrogen and oxygen emission cross sections for 
electron impact on H$_2$O  based on the 
measurements from 
\cite{Mohlmann78} and \cite{Morgan74}, respectively.
We have noticed that there is a discrepancy, which will be 
discussed later, between \cite{Itikawah2o}  
and \cite{Shirai01} cross sections  for  H~{\scriptsize{I}} 
Lyman-$\alpha$ 
emission.
We have used \cite{Shirai01} cross section
to calculate H~{\scriptsize{I}} Lyman-$\alpha$\ emission via electron 
impact on H$_2$O in the model.

 {The recent measured various  atomic oxygen and hydrogen emission 
 cross sections via  electron impact dissociative excitation of 
 H$_2$O  at 100 and 200 eV, are from \cite{Makarov04}.} The cross 
 section 
ratio of H~{\scriptsize{I}} Lyman-$\alpha$/Lyman-$\beta$ 
(H~{\scriptsize{I}} Lyman-$\alpha$/Lyman-$\gamma$)  from 
 {these measurements} at electron energy of 200 eV is 8 (30.5). 
In order to construct the cross section 
for
H~{\scriptsize{I}} Lyman-$\beta$\ emission via electron impact on 
H$_2$O, 
\cite{Feldman15} 
reduced \cite{Itikawah2o} suggested H~{\scriptsize{I}} 
Lyman-$\alpha$\ 
cross section by a factor 8 \citep{Chaufray17}. To 
incorporate 
H~{\scriptsize{I}} Lyman-$\beta$\ and H~{\scriptsize{I}} 
Lyman-$\gamma$\ emission 
processes
via electron impact on H$_2$O, we have also reduced \cite{Shirai01}  
H~{\scriptsize{I}} Ly-$\alpha$ cross section by the corresponding 
cross section ratios  {from the measurements} of \cite{Makarov04}, 
 {that } are determined at 200 eV.

\subsubsection{Cross sections for  atomic oxygen 
	emissions:  O~{\scriptsize{I}} 1152, 1304, and 1356 \AA}  
 {\cite{Wu88} have measured different atomic hydrogen and oxygen 
 emission cross sections by dissociating H$_2$O using the photons in 
 the energy 
range 15--20.5 eV.} They
have also extensively discussed all the possible  
H$_2$O photodissociative channels and also emissions 
from the dissociated species \citep[See Table 1 
of][]{Wu88}. Based on the sensitivity of  {their} instrument, they 
set an upper limit of the cross section for all 
the unobserved emissions as  5 $\times$ 10$^{-21}$ cm$^{2}$ 
in the measured wavelength region. The suggested quantum yield 
of all the unobserved  emissions was $<$ 2.2 $\times$ 10$^{-4}$
in the wavelength region 600--825 \AA. We have used this yield to estimate 
 {the O~{\scriptsize{I}} 1152, 1304, 
and 1356 \AA\ emission cross sections for photodissociative excitation of 
H$_2$O}.

The  {O~{\scriptsize{I}} 1304 \AA\ emission cross sections for 
photodissociation
of CO and CO$_2$ }are taken from  
\cite{Wu79a}.  We could not find the directly measured 
 { O~{\scriptsize{I}} 1152, 1356~\AA\ emission cross sections for 
photodissociative excitation of CO and CO$_2$} in the  
literature. However, \cite{Lee75} measured fluorescence cross 
section for photodissociative fragments of CO and CO$_2$ in the
wavelength region 175--800 \AA\ and the observed
total emission cross section is attributed to various 
onsets of 
atomic carbon and oxygen emissions. 
Due to the proximity in the threshold energies,
we have assumed that  {the cross section for O~{\scriptsize{I}}
 1304~\AA\ emission via photodissociative excitation CO and CO$_2$}
 is same as that of O~{\scriptsize{I}} 1152 and 1356~\AA. 
The effect of our assumption on the calculated total oxygen 
emission intensities will be discussed in 
Section~\ref{sec:effect-inputs}.

\cite{Carlson74} measured  { O~{\scriptsize{I}} 7774~\AA\ 
($^5$P $\rightarrow$ $^5$S) and 1302~\AA\  resonant multiplet
($^3$S $\rightarrow$ $^3$P) emission cross 
sections by  photodissociative excitation of O$_2$ in the} wavelength regions 
 730--780 \AA\ and 650--850 \AA, respectively.  
 The  O~{\scriptsize{I}} 7774 \AA\ 
emission leads to the formation of excited 
O($^5$S) which subsequently decay to ground O($^3$P) state
by producing photons at wavelength 1356 \AA, provided 
if it does not collisionally quench by surrounding cometary 
species. \cite{Zhou14} measured the photodissociation cross sections
of O$_2$\ producing O($^3$S) and O($^5$S) states, which are 
the excited states of  {O~{\scriptsize{I}}} 1302 \AA\ and 1356 \AA\ 
 {emissions}, 
respectively, in the energy range 14.64--15.20 eV 
(815--847 \AA).
They found that the production of 
O($^5$S) is higher than that of O($^3$S) in the
photodissociation of O$_2$ with a mean dissociative 
excitation cross section ratio of 1.5.  
To incorporate  {O~{\scriptsize{I}} 1356 \AA\ emission emission
via photodissociation of O$_2$} in 
our model, we have multiplied \cite{Carlson74} measured  
O~{\scriptsize{I}} 1302 \AA\ cross sections by a factor 1.5. 
The mean electron impact cross section ratio of O$_2$
for the production of O~{\scriptsize{I}} 1152 to O~{\scriptsize{I}} 
1304 \AA\  is about 2.8 \citep{Kanik03}. 
We have assumed that this branching ratio is same for 
the photodissociation of O$_2$. So we have divided \cite{Carlson74} 
measured  O~{\scriptsize{I}} 1304 \AA\  cross section 
by a factor 2.8 to calculate the O~{\scriptsize{I}} 
1152 \AA\ emission intensity via O$_2$ photodissociation.

We have used \cite{Shirai01} suggested  { O~{\scriptsize{I}} 1304 
\AA\ emission cross section for electron impact on H$_2$O}. Based on the 
\cite{Makarov04} cross sections 
measurements on H$_2$O at 200 eV (100 eV), we have determined 
the electron impact dissociative excitation 
cross section ratios of H~{\scriptsize{I}} 
Lyman-$\alpha$/O~{\scriptsize{I}} 1152 (O~{\scriptsize{I}} 
1304/1356) as 29 (1.5). 
These  ratios have been used to reduce \cite{Shirai01} 
H~{\scriptsize{I}} Lyman-$\alpha$\ and O~{\scriptsize{I}} 
1304 cross sections, respectively, to incorporate
electron impact on H$_2$O producing O~{\scriptsize{I}} 1152 
and O~{\scriptsize{I}} 1356 \AA\ emissions in the model. 
The  {O~{\scriptsize{I}} 7774 \AA\ emission
cross section by electron impact  of H$_2$O }is taken 
from \cite{Beenakker74}.

The  {O~{\scriptsize{I}} 1356 \AA\ emission cross section 
by electron impact  of CO$_2$} is taken from 
\cite{Ajello71-co2}. We have considered the
 {cross sections for O~{\scriptsize{I}} 1304 \AA\ emission
by electron impact of CO and CO$_2$} from 
\cite{Ajello71-co} and \cite{Itikawaco2}, respectively. 
\cite{Ajello71-co} also determined emission cross section 
ratio of O~{\scriptsize{I}}~1356/1304 for electron impact on 
CO at 100 eV as 0.85.  So we have multiplied 
\cite{Ajello71-co} O~{\scriptsize{I}} 1304 cross section 
with this factor to obtain O~{\scriptsize{I}} 1356 emission 
cross section for electron impact on CO.

\cite{Kanik95} and \cite{Kanik93} determined the
 {O~{\scriptsize{I}} 1152 \AA\ emission cross section by
electron impact of CO and CO$_2$}  at  200 eV as 
about 3.5 $\times$ 10$^{-19}$ cm$^2$, respectively.
This  {cross section} value is higher (lower) by a factor 2 to the 
\cite{Ajello71-co}  \citep{Itikawaco2} measured 
O~{\scriptsize{I}} 1304 emission cross section.
 So we have scaled  the \cite{Ajello71-co} 
and \cite{Itikawaco2} O~{\scriptsize{I}} 1304 \AA\ emission 
cross sections with the corresponding scaling factor to 
incorporate O~{\scriptsize{I}} 1152 \AA\ emission via 
electron impact on CO and CO$_2$, respectively.
 {The O~{\scriptsize{I}} 1152, 1304 
and 1356 \AA\ emission cross sections by electron impact dissociative excitation
of O$_2$} are taken from \cite{Kanik03}.

\subsubsection{Cross sections for  atomic carbon 
	emissions : C~{\scriptsize{I}}  1561 and 1657 \AA} 
 {The C~{\scriptsize{I}} 1561 and 1657 
\AA\ emission cross sections by photodissociation of
CO and CO$_2$} are considered from \cite{Wu81} and 
\cite{Wu78}, respectively, and  {these emission cross sections 
for the electron impact on CO$_2$} are taken 
from \cite{Ajello71-co2}. 
We have followed the approach of \cite{Bhardwaj96} to 
consider the  {C~{\scriptsize{I}} 1561 and 
1657 \AA\  emission cross sections for the electron impact of CO,} 
which earlier has been used by \cite{Paxton85} to calculate 
these emission intensities  on Venus.

\section{Description of model calculations} 
\label{model-des}
The neutral atmospheric density profiles of cometary 
species are calculated using Haser's formula, which assumes 
spherical expansion of volatiles into the space 
\citep{Haser57}. Neutral gas expansion velocity from the nucleus is taken 
as 1 km/s. The solar photons, in the 
wavelength region 5--1900 \AA,  have been degraded in 
the cometary neutral atmosphere.  {The primary 
photoelectron energy spectrum $Q(E, r, \theta)$ is 
calculated using the following expression.}
\begin{equation}
Q(E, r, \theta) = \sum_i 
\int_{\lambda_{min}}^{\lambda_{max}}\ n_i(r)\ 
\sigma_i^I(\lambda) \ 
\ I_\infty(\lambda)\  exp[-\tau(r, \theta, \lambda)]\  
d\lambda
\end{equation}
where 

\begin{equation}
 \tau(r, \theta, \lambda) = \sum_i \sigma_i^A(\lambda) \ 
 sec\theta \int_{r_o}^{r_\infty} n_i (r')\ dr'
\end{equation}

Here $\sigma_i^A(\lambda)$ and 
$\sigma_i^I(\lambda)$ are the absorption and ionization 
cross sections (cm$^2$)  at wavelength $\lambda$, 
respectively, and $n_i(r)$ is the 
number 
density (cm$^{-3}$) at radial distance $r$ of the $ith$ 
neutral species. 
$I_\infty(\lambda)$ is unattenuated solar photon flux 
(cm$^{-2}$ s$^{-1}$) at the 
top of atmosphere. $\tau(r, 
\theta, \lambda)$ is the optical depth of the medium at 
solar zenith angle $\theta$.  $r_o$ and $r_\infty$ are 
radius of comet (taken as 2 km) and top of the atmosphere 
(taken as 10$^5$~km), respectively. We have used Analytical 
Yield Spectrum (AYS) approach to calculate steady state 
photoelectron flux in 
the cometary coma. The AYS method  of degrading electrons 
in the neutral atmosphere can be  explained briefly in the 
following manner. Mono-energetic electrons incident along 
the  {radial axis} in an infinite medium 
are degraded in a collision-by-collision manner using the 
Monte Carlo technique. The energy and position of the 
primary electron and its secondary and tertiary are 
recorded at the instant of an inelastic collision. The 
total number 
of inelastic events in the spatial and energy bins, after 
the incident electron and all its secondaries and 
tertiaries have been completely degraded, is used to 
generate numerical yield spectra. These yield spectra 
contain the information about the electron 
degradation process and can be employed to calculate the 
yield for any inelastic event. The numerical yield spectra, generated 
in this way, are fitted by an analytical expression
to provide AYS. This yield 
spectrum can be used to calculate the steady state 
photoelectron flux. More details of the AYS approach and 
the method of photoelectron computation are given in 
several previous papers  
\citep{Singhal84,Bhardwaj90,Bhardwaj96, Singhal91, 
Bhardwaj99b, Bhardwaj03, Bhardwaj99a, Haider05,  
Bhardwaj09,Bhardwaj12}. The photoelectrons are degraded
based on the local energy deposition approximation, hence 
this method of calculation of suprathermal electron flux  
is valid  within the inner coma where the strong collisional
coupling between  neutrals and electrons exist. When 
gas production rate of comet is low (about 10$^{26}$ or less), 
this assumption is not valid since electrons can not 
transfer energy efficiently to the surrounding neutrals 
due to low gas density.  This model does not account for 
interaction between cometary nucleus surface generated 
photoelectrons and solar wind electrons with the neutral 
species of coma.

Volume emission rates for different excited species 
are calculated by using degraded photon and suprathermal 
electron flux profiles and corresponding excitation cross 
sections.  The emission intensities of various spectral 
lines, as a function of nucleocentric projected distance, 
are  obtained by integrating the volume emission 
rates along the Alice line of sight and converting 
brightness in Raleigh (1 R = 10$^6$/4$\pi$ photons 
cm$^{-2}$ s$^{-1}$ sr$^{-1}$). When the 
Alice spectrometer is pointed towards 67P/C-G 
nucleus, the nadir intensities are calculated by 
integrating the volume emission rates from surface of the 
nucleus to spacecraft distance  {on the illuminated part of the line of 
sight.}
 
To incorporate the resonance
fluorescence mechanism for the allowed atomic transitions, 
neutral density profiles of atomic oxygen and carbon 
are calculated by solving the continuity equation as described 
by \cite{Bhardwaj96}.  Atomic hydrogen number density 
profile is calculated using Haser's formula. There is a
discrepancy (about an order of magnitude) between Haser's 
calculated and the direct simulation of Monte-Carlo (DSMC) 
modelled atomic hydrogen number density profiles and its 
impact on the calculated emission intensity will be discussed
in  {Section~\ref{results:mden}.} 
Resonance fluorescence efficiencies for H~{\scriptsize{I}}
 Lyman-$\alpha$, 
Lyman-$\beta$, Lyman-$\gamma$, O~{\scriptsize{I}} 1304, 
C~{\scriptsize{I}} 1561 and 1657 \AA\ are taken as 2.88 
$\times$ 10$^{-3}$, 3.84 $\times$ 10$^{-6}$,
 7.33 $\times$ 10$^{-7}$,  2.22 $\times$ 10$^{-6}$, 
 3.08 $\times$ 10$^{-5}$, and
1.1 $\times$ 10$^{-5}$ s$^{-1}$, respectively, at 1 AU 
\citep{Gladstone10, Meier95, Woods86} and these values are scaled 
to the corresponding heliocentric distances of comet 67P/C-G.

Most of the emissions considered in this work are 
due to spontaneous decay and the lifetime of the excited 
species is about a few nanoseconds, except for  
O~{\scriptsize{I}} 1356 \AA\ (180 $\mu$s). Hence, the 
collisional quenching is not a significant loss mechanism 
in removing the excited H~{\scriptsize{I}}, 
O~{\scriptsize{I}}, and  C~{\scriptsize{I}} states compared to  
radiative decay. Owing to a small 
radiative lifetime ($<<$ 1 s), the excited species cannot 
travel to a large radial distance in the cometary coma from the 
origin of production before emitting photons via radiative 
decay, hence, the radial transport of these species can be 
neglected.  {When a comet is having gas production rate about 10$^{28}$ 
s$^{-1}$, the 1/r$^2$ dependence of the neutral number densities
reflects the optically thin cometary coma for the radial distances 
above 10 km  \citep[][]{Vigren13,Bhardwaj03}.
Hence, the emission intensities are calculated assuming optically
thin condition in the coma, which is a reasonably good 
approximation for a low water production comet 
of 67P/C-G ($<$10$^{27}$ s$^{-1}$).} 	

\section{Results}
\label{sec:results}

\begin{figure}[htb]
	\centering
	\includegraphics[width=\columnwidth]{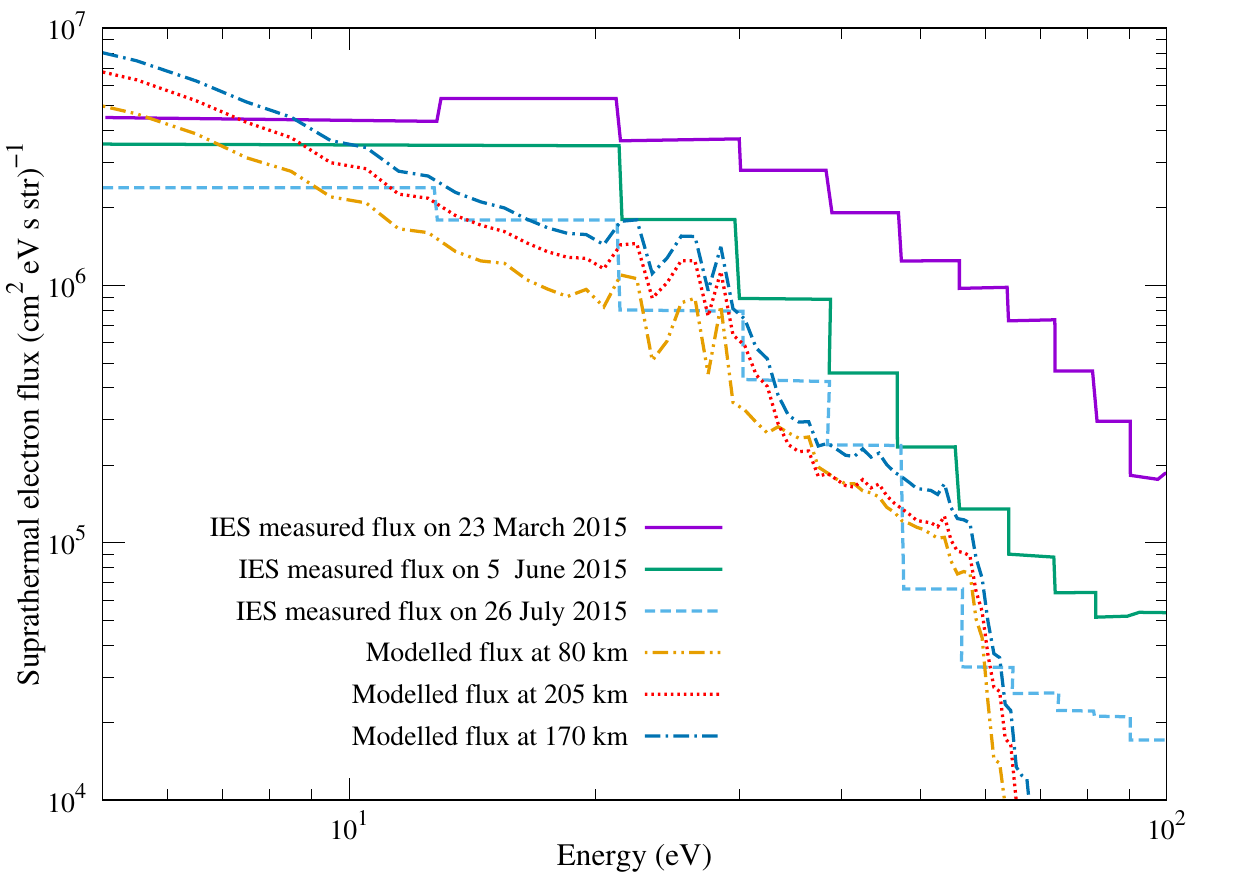}
	\caption{Comparison between model calculated 
	and daily averaged RPC-IES suprathermal electron fluxes 
	 observed on 23 March (solid line), 5 June (dashed), 
	 and  26 July 2015 (thick dashed)	\citep{Madanian16, 
	 Madanian17}  when Rosetta was at around 80, 205, and 
	 170 km radial distances from the surface of the comet. 
	The model calculated fluxes on these respective days 
	are plotted with  dash-double-dotted, fine dotted, 
	dash-dotted and  curves. We have taken  respective 
	total gas production rates on these days, when comet 
	67P/C-G when was at 2, 1.5 and 1.26 AU heliocentric 
	distances,  as  2 $\times$ 10$^{26}$, 3 $\times$ 
	10$^{27}$  and 1 $\times$ 10$^{28}$ s$^{-1}$ from 
	\cite{Hansen16}.} 
    \label{fig:cmp_eflx}  
\end{figure}

The modelled suprathermal electron fluxes along with 
the daily-averaged RPC-IES measurements made on 2015 
March  23, June 5, and July  26 \citep{Madanian16, 
Madanian17} are shown in Figure~\ref{fig:cmp_eflx}.
During these observations, Rosetta was at 2.0, 1.5 and 
1.26 AU heliocentric distances and orbiting at radial 
distances of 80, 205, and  170 km away from the nucleus, 
respectively. For the calculation of photoelectron flux 
on these days, we have taken the respective total gas 
production rates as 2 $\times$ 10$^{26}$, 3 $\times$ 
10$^{27}$, and 1 $\times$ 10$^{28}$ s$^{-1}$ from Rosina-DFMS 
measurements \citep{Hansen16}. The peaks between 20 and 30 eV 
in the modelled suprathermal flux are mainly due to 
photoionization of H$_2$O by solar He~{\scriptsize{II}}
304 \AA\ photons, while the fall above 60 eV is due 
to a sharp decrease (about two orders of magnitude) 
in solar flux in the wavelength region 125 -- 175 \AA.

\begin{figure}[htb]
	\centering
	\includegraphics[width=\columnwidth]{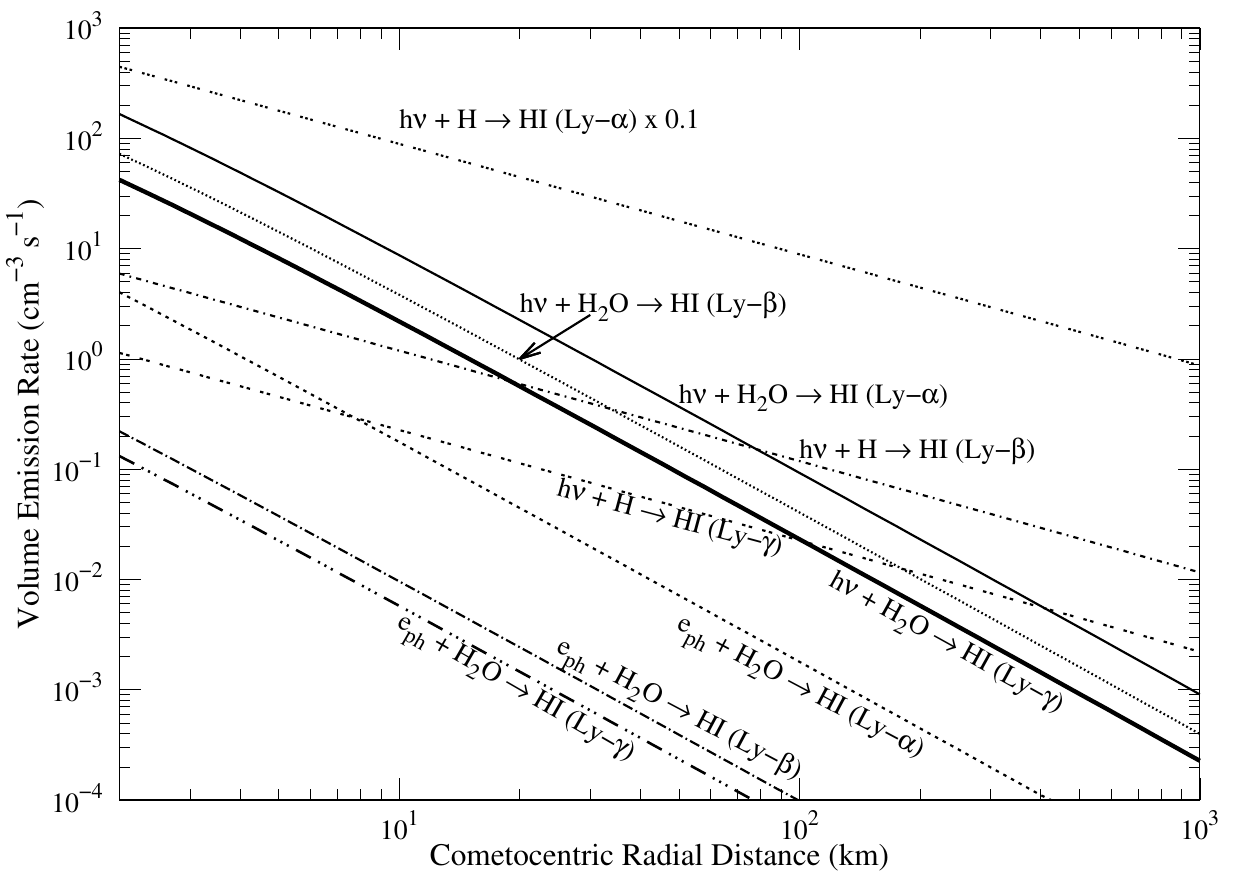}
	\caption{The model calculated volume emission rates for 
		atomic hydrogen Lyman emissions, via 
		photon,	photoelectron impact dissociative 
		excitation of H$_2$O and resonance fluorescence of 
		atomic hydrogen, on comet 67P/C-G\ at heliocentric 
		distance 1.56 AU.  The calculations are done for 
		water production rate 2	$\times$ 10$^{27}$ 
		s$^{-1}$ and with relative abundances of 3, 2, and 
		4\% of CO$_2$, CO, and O$_2$\ with respect to 	
		water, respectively. Volume emission rate profile 
		of H~{\scriptsize{I}} Lyman-$\alpha$ via solar 
		resonance fluorescence is multiplied by a factor 
		0.1. h$\nu$ is solar photon and e$_{ph}$ is 
		photoelectron.} 
	\label{fig:hydro-emis}  
\end{figure}

The calculated volume emission rate profiles for 
H~{\scriptsize{I}} Lyman emissions (Ly-$\alpha$, 
Ly-$\beta$, \& Ly-$\gamma$) via different 
excitation mechanisms, when the comet was at heliocentric distance of 
1.56 
AU,  are 
shown in Figure~\ref{fig:hydro-emis}.
This calculation shows that the solar resonance 
fluorescence is the dominant excitation source for 
H~{\scriptsize{I}} Ly-$\alpha$  emission in the entire coma 
compared to the dissociative excitation of H$_2$O via photons 
and suprathermal electrons.  For  H~{\scriptsize{I}} Ly-$\beta$ 
(Ly-$\gamma$) emission, the photodissociation of H$_2$O is the 
major source for the radial distances below 30 (100) km above 
which solar resonance fluorescence is the significant excitation 
mechanism. This calculation also shows that the production of
H~{\scriptsize{I}} emission lines via electron impact 
excitation of H$_2$O is an order of magnitude lower compared 
to the photodissociative excitation of H$_2$O.

\begin{figure}[htb]
	\centering
	\includegraphics[width=\columnwidth]{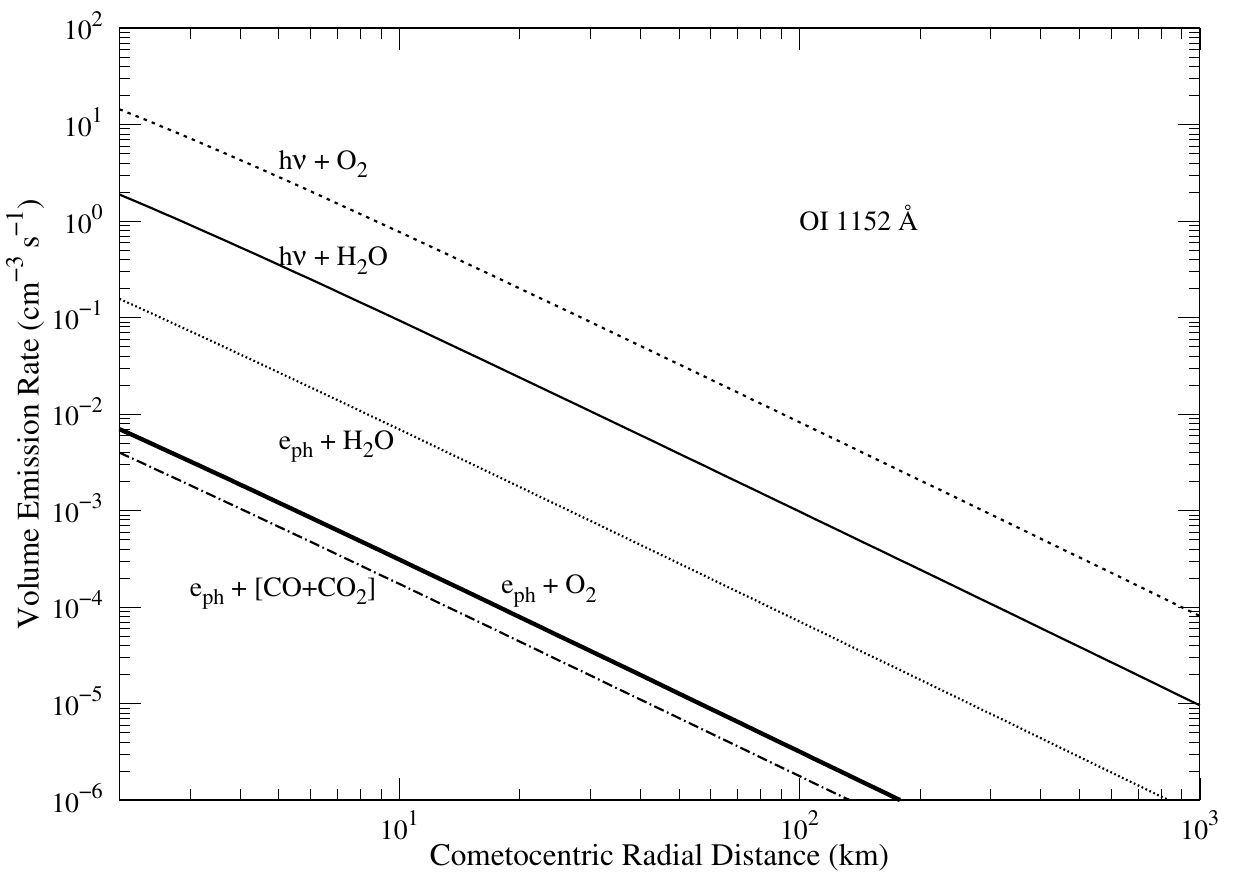}	
	\caption{The model calculated volume emission rates for 
		atomic  oxygen 1152 \AA\ emission as a
		function of cometocentric radial distance on comet 
		67P/C-G at heliocentric distance of 1.56 AU for the 
		same composition as mentioned in 
		Figure~\ref{fig:hydro-emis}. Volume emissions 
		rates of electron impact on CO and CO$_2$ 
		are added due their proximity. h$\nu$ is 
		solar photon and e$_{ph}$ is photoelectron. }
	\label{fig:emis-inten-oa}	
\end{figure}

\begin{figure}[htb]	
	\centering	
	\includegraphics[width=\columnwidth]{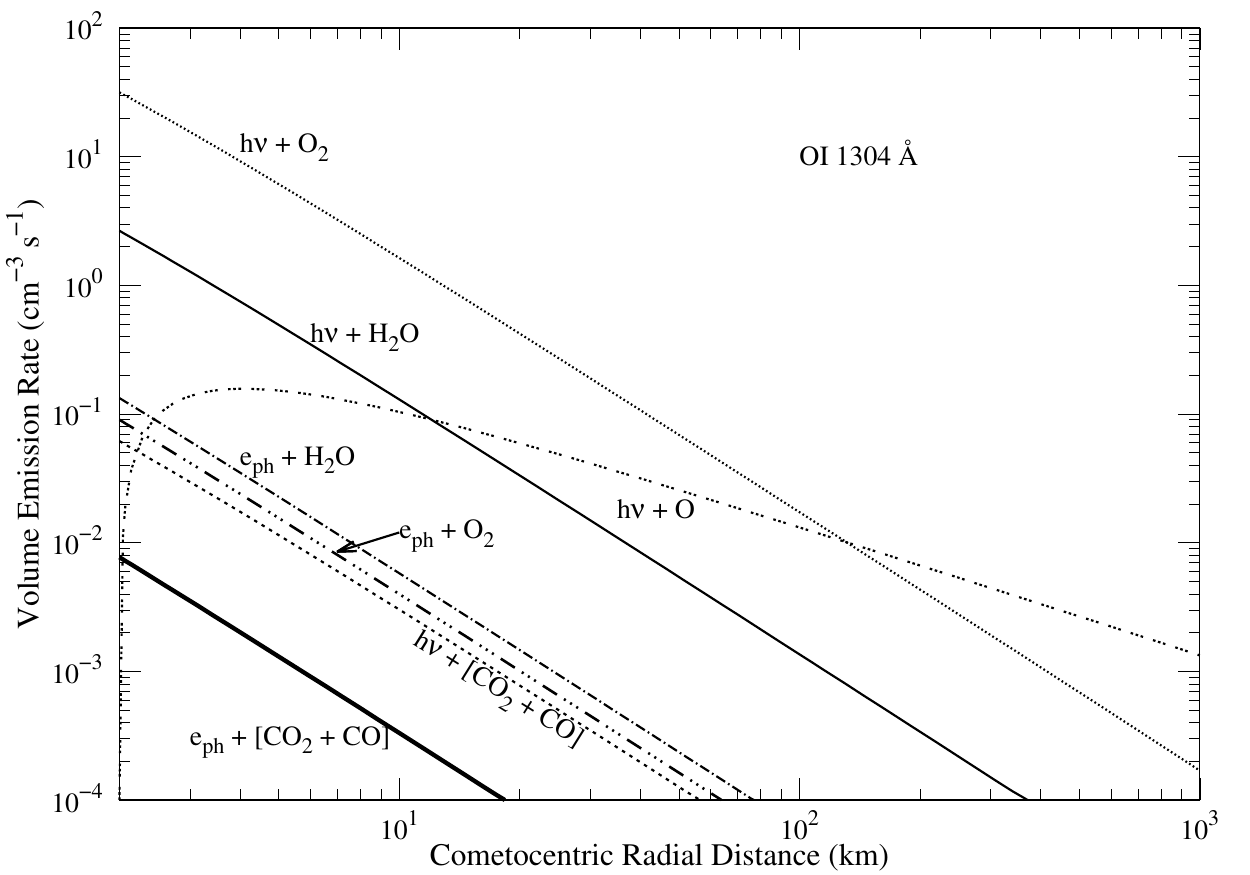}
	\caption{The model calculated volume emission rates for 
		atomic  oxygen 1304 \AA\ emission as a
		function of cometocentric radial distance on 
		comet 
		67P/C-G at heliocentric distance of 1.56 AU for the 
		same composition as mentioned in 
		Figure~\ref{fig:hydro-emis}. Dissociative 
		excitation of CO$_2$ and CO via photons and  
		electrons are shown single curves due 
		to their proximity. 
		h$\nu$ is solar photon and e$_{ph}$ 
		is photoelectron.}
	\label{fig:emis-inten-ob}	
\end{figure}

\begin{figure}	
	\centering		
\includegraphics[width=\columnwidth]{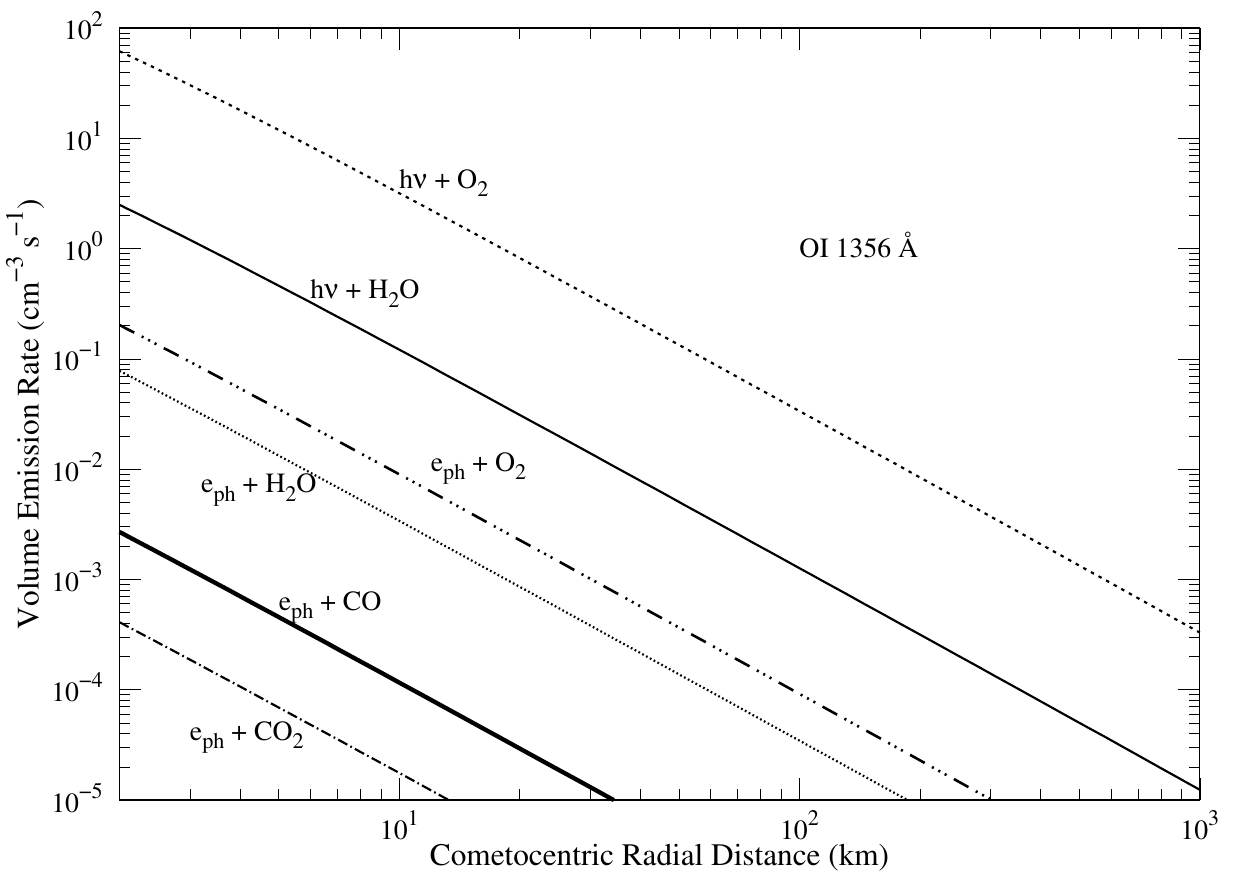}
	\caption{The model calculated volume emission rates for 
		O~{\scriptsize{I}} 1356 \AA\ via photon and 
		photoelectron impact dissociation 
		processes on comet 67P/C-G\ at heliocentric 
		distance of 1.56 AU. Input conditions are same as 
		that are mentioned in Figure~\ref{fig:hydro-emis}.  
		h$\nu$ is solar photon and e$_{ph}$ is 
		photoelectron.}
	\label{fig:emis-inten-oc}
\end{figure}

The modelled volume emission rate profiles of 
O~{\scriptsize{I}} 1152, 1304 and 1356 \AA\ emissions, when the comet 
was 
at heliocentric distance of 1.56 AU, 
are presented in Figures~\ref{fig:emis-inten-oa}, 
\ref{fig:emis-inten-ob}, and \ref{fig:emis-inten-oc}, 
respectively. It is found that photodissociative excitation 
of molecular oxygen is the primary source of 
O~{\scriptsize{I}} emissions followed by H$_2$O 
photodissociation.  The contribution of total electron 
impact on O-bearing neutrals  is  more than an order of 
magnitude smaller compared to photodissociative excitation. 
Above radial distances of 150 km, resonance fluorescence 
is the dominant source of O~{\scriptsize{I}} 1304 \AA\ 
emission (see Fig.~\ref{fig:emis-inten-ob}).

\begin{figure}		
	\centering	
	\includegraphics[width=\columnwidth]{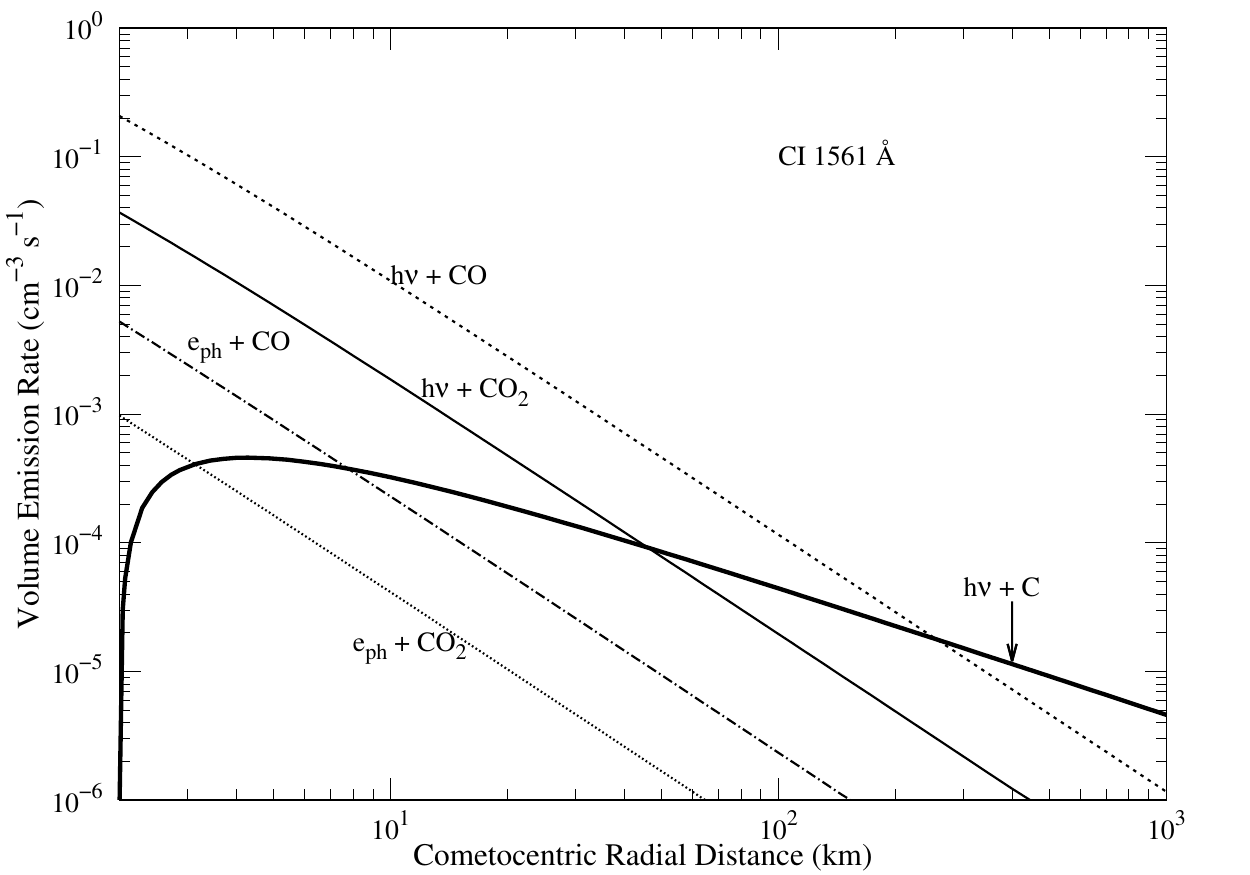}
	\includegraphics[width=\columnwidth]{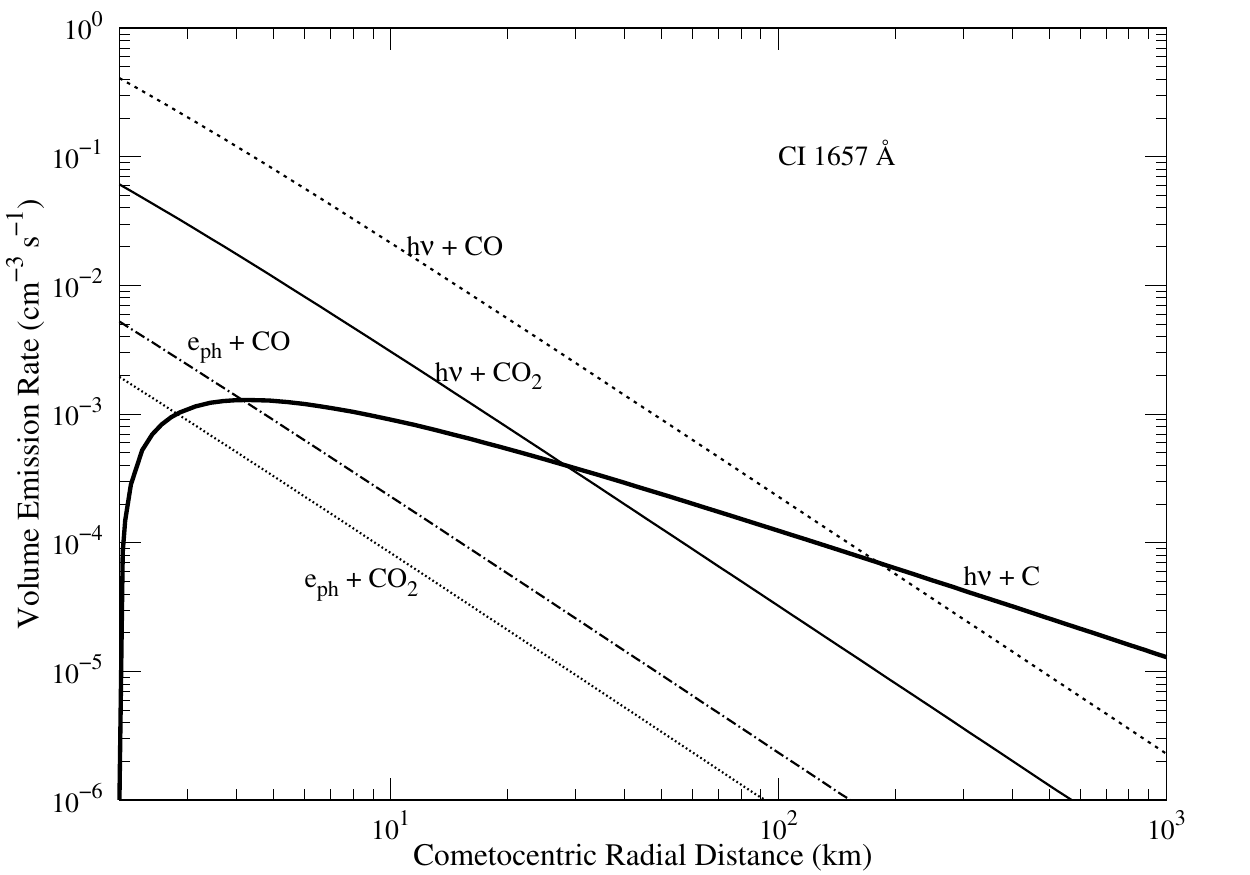}
	
	\caption{The model calculated volume excitation 
	rates 		
	for atomic carbon emissions at wavelengths 1561 (top 
	panel) and 1657 \AA\ (bottom panel) via different 
	excitation mechanisms on comet 67P/C-G\ at heliocentric 
	distance of 1.56 AU. Input conditions are same as that 
	are mentioned in Figure~\ref{fig:hydro-emis}. h$\nu$ is 
	solar photon and e$_{ph}$ is photoelectron.}
	\label{fig:carb-emis-a} 	
\end{figure}

Our model calculated volume emission rate profiles 
for C~{\scriptsize{I}}  emissions
via different excitation mechanisms are shown in  Figure 
\ref{fig:carb-emis-a}. These calculations show that 
photodissociative excitation of CO is the most dominant 
source in producing C~{\scriptsize{I}} 1561 (1657) \AA\ emission 
followed by photodissociative excitation of CO$_2$ for radial 
distances below 50 (30) km. The resonance fluorescence of atomic 
carbon is the dominant source for radial distances larger than 
300 km. The calculated volume emission rates for electron impact 
on CO and CO$_2$ are found to be smaller by an order of magnitude 
compared to respective photodissociative excitation rates.

\begin{figure}[htb]
	\centering
	\includegraphics[width=\columnwidth]{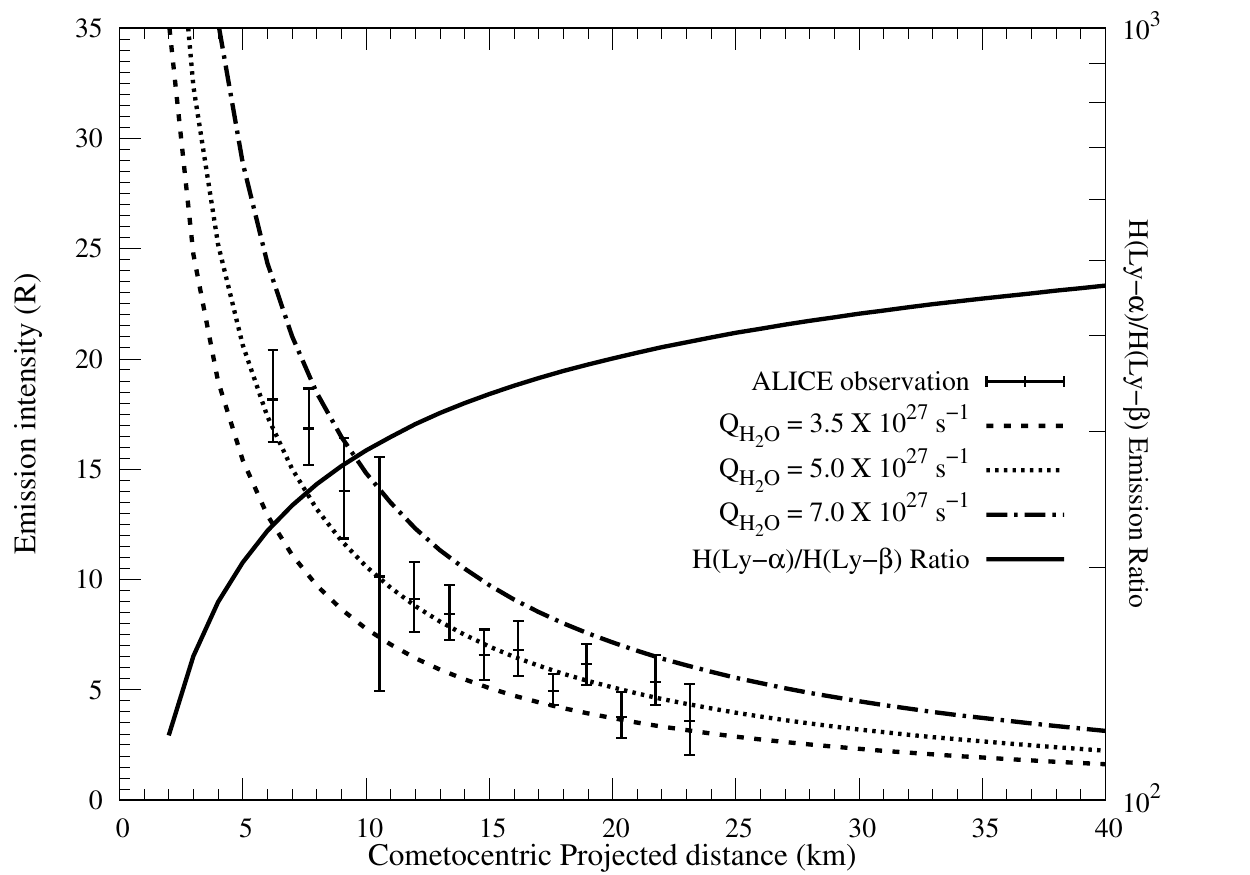} 		
	\caption{The model calculated H~{\scriptsize{I}} 
		Lyman-$\beta$ emission intensity 
		as a function of cometocentric projected distance 
		on 
		comet 67P/C-G, when it was at 1.56 AU from the Sun,
		for three different water production rates (Q$_{H_2O}$). 
		 {Alice} observed H~{\scriptsize{I}} 
		Lyman-$\beta$ 
		emission intensity on 25 May 2015, from 
		\cite{Chaufray17}, is plotted with 
		vertical error bars. 
		The calculations are done by varying H$_2$O 
		production 
		rate between  3.5 and 7 $\times$ 10$^{27}$ s$^{-1}$ 
		and for relative abundances of 2.5, 2, and 4\% of 
		CO$_2$, CO, and O$_2$\ with respect to water, 
		respectively. Emission intensity is calculated in 
		Rayleigh (1 R = 10$^6$/4$\pi$ photons cm$^{-2}$ 
		s$^{-1}$).  {The modelled H~{\scriptsize{I}} 
			Lyman-$\alpha$/H~{\scriptsize{I}} 
			Lyman-$\beta$ emission intensity ratio is plotted on right y-axis.}}
	\label{fig:emis-inten}
\end{figure}

We have calculated the limb intensity 
profiles for H~{\scriptsize{I}} Lyman-$\beta$ emission as a function 
of cometocentric projected distance for the Alice observational 
conditions on 2015 May 25, which is shown in 
Figure~\ref{fig:emis-inten}. By varying the H$_2$O production rate 
within the ROSINA measured limits, i.e., from 3.5 to 7 $\times$ 10$^{27}$ 
s$^{-1}$, we find a good correlation between the modelled  and Alice observed
 H~{\scriptsize{I}} Lyman-$\beta$  intensity profiles \citep{Chaufray17}. 
  {In this case, we have also calculated  H~{\scriptsize{I}} 
  Lyman-$\alpha$/H~{\scriptsize{I}} 
Lyman-$\beta$ intensity ratio as a function of cometocentric projected 
distance. By varying the H$_2$O production rate, as mentioned before,
we could not find a significant change ($<$5\%) in the modelled  
H~{\scriptsize{I}} emission ratio.}

\begin{figure}
	\centering	
	\includegraphics[width=\columnwidth]{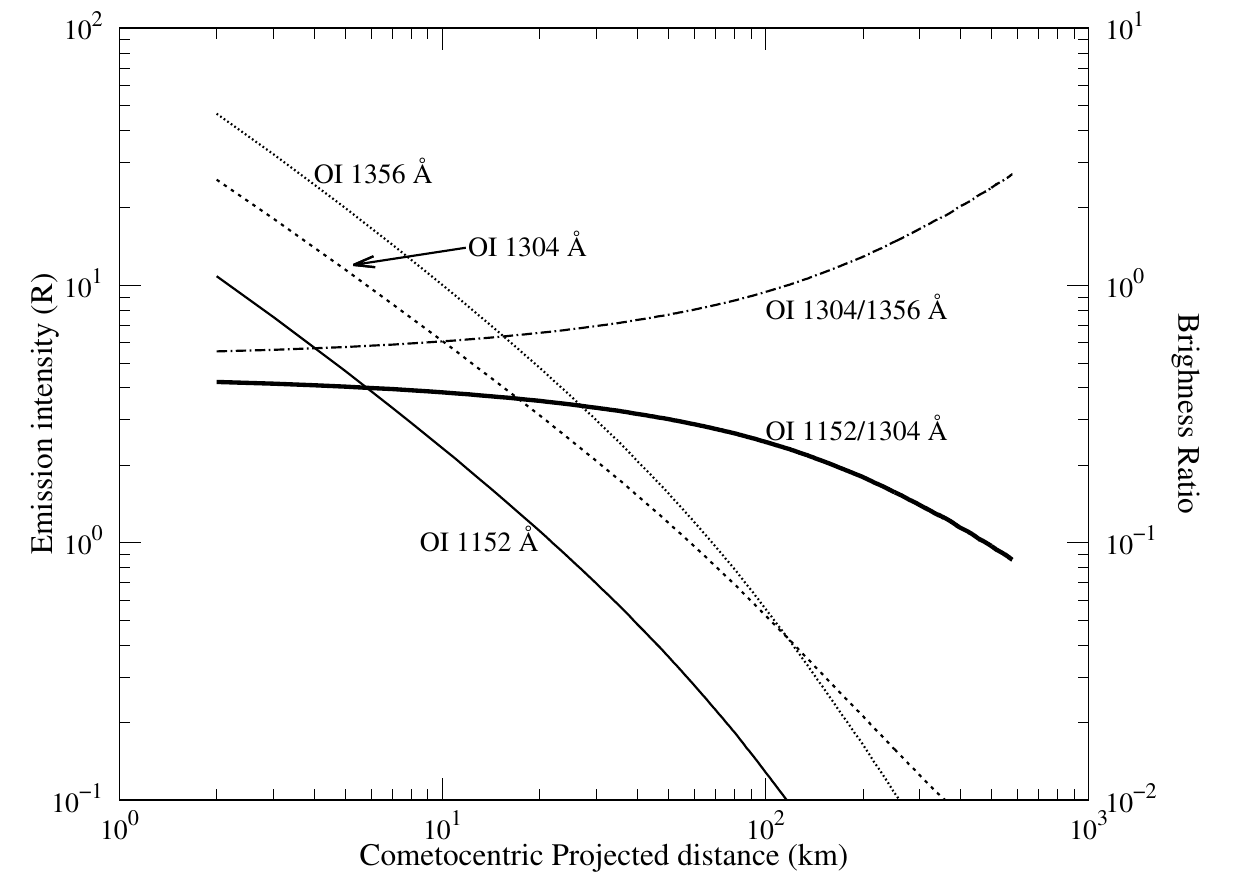}	
    \includegraphics[width=\columnwidth]{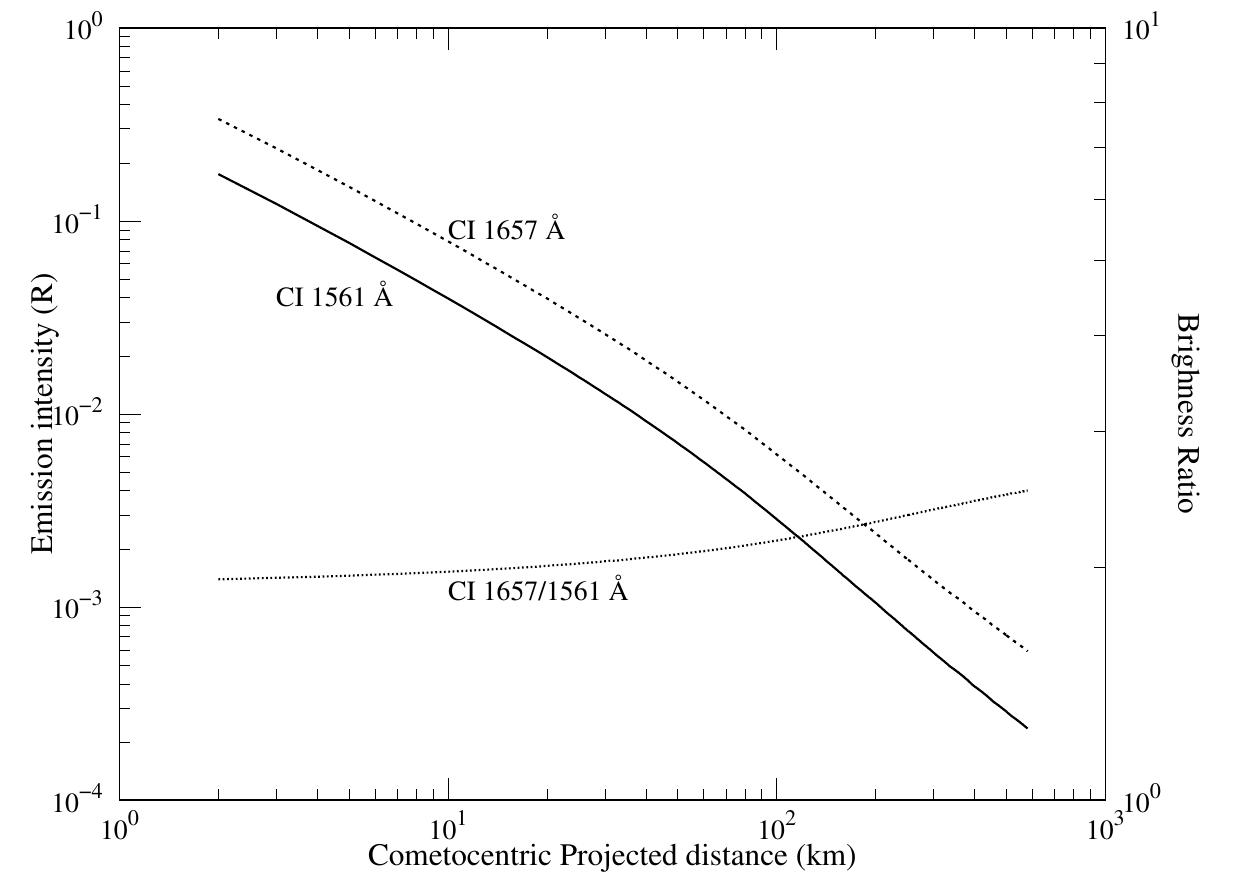}
	\caption{The model calculated emission intensities for 
		atomic oxygen (top panel), and carbon (bottom 
		panel) emissions as a function of cometocentric 
		projected distance on comet 67P/C-G at a 
		heliocentric distance of 1.56 AU. The oxygen 
		1304/1356 \AA\ and 
		O~{\scriptsize{I}} 1152/1304 \AA\ emission 
		ratios are plotted on right y-axis in the 
		top panel. Similarly C~{\scriptsize{I}} 1657/1561 
		\AA\ emission ratio is plotted on right y-axis in 
		the bottom panel. Input conditions are same as 
		that are mentioned in Figure~\ref{fig:hydro-emis}.	}
    	\label{fig:emis-oxy-carb}
\end{figure}

By considering various dissociative excitation channels 
of carbon and oxygen bearing species, the calculated limb
intensity profiles for O~{\scriptsize{I}} (1152, 1304, 1356 \AA), 
C~{\scriptsize{I}} (1561 and 1657 \AA) emissions
 are shown in 
Figure~\ref{fig:emis-oxy-carb}.
In this figure, we have also presented the 
O~{\scriptsize{I}} and  C~{\scriptsize{I}} emission brightness 
ratio as a function of projected distances on the right 
Y-axis. The modelled O~{\scriptsize{I}} 1304/1356 
\AA\ (O~{\scriptsize{I}} 1152/1304 \AA) brightness ratios are found to be 
increasing (decreasing) with increasing projected 
distance, whereas the calculated C~{\scriptsize{I}} 1657/1561 \AA\ 
emission ratio is around 2 for the projected distances less 
than 1000 km.

We have made calculations on 2015 March 29, when the comet was at 
1.99 AU from the Sun, during which the Alice spectrometer was pointed 
towards nucleus and the solar phase angle was about 
80$^\circ$. By varying the chemical composition in the cometary coma, 
 {the modelled and observed H~{\scriptsize{I}} (Ly-$\alpha$), 
O~{\scriptsize{I}} (1304 and 1356 \AA) and C~{\scriptsize{I}} (1657 \AA) 
intensities are}  presented in 
Table~\ref{tab:results}.  In this case, since our modelled 
photoelectron flux is
smaller by a factor 5 or more  compared 
to the RPC-IES observation (see Figure~\ref{fig:cmp_eflx}), we have used 
RPC-IES measured photoelectron flux by assuming  {its magnitude vary as 
a function of inverse of the distance to the nucleus in the inner coma 
\citep{Edberg15}}. Even by 
using 
RPC-IES measured photoelectron flux in the model, we find 
photodissociative excitation reactions play a significant role 
(a factor of 5 or more) than the electron impact excitation reactions.
At this heliocentric distance, H$_2$O 
production rate  {($Q_{H_2O}$)} is found to vary in the range 2--12.5 
$\times$ 10$^{26}$ s$^{-1}$ \citep[see Table~\ref{tab:neu}
and][]{Hansen16}. When we use  {$Q_{H_2O}$ as  2 
	$\times$ 10$^{26}$ s$^{-1}$}, the modelled
emission H {\scriptsize{I}} Lyman-$\beta$ intensity is consistent 
with the Alice observation, whereas the calculated 
O {\scriptsize{I}} emission intensities are smaller by a factor of about
4. In this 
case, the calculated C~{\scriptsize{I}} emission intensity is smaller 
by more than an order of magnitude.  Considering  {$Q_{H_2O}$ as}  12.5 
$\times$ 10$^{26}$ s$^{-1}$ and for the 
previously mentioned relative abundances of other 
species, the model determined  
O~{\scriptsize{I}} intensities are close to 
Alice observation and the calculated H~{\scriptsize{I}} 
(C~{\scriptsize{I}}) 
nadir emission intensities are higher by a factor 5 (lower by an order of 
magnitude)  {compared to the observation.}  {In this case, contrary 
to the model calculations, the 
	observed O~{\scriptsize{I}} 1304 \AA\ emission intensity is higher than 
	that of 
	O~{\scriptsize{I}} 1356 \AA. }
 {By increasing the} gas production rate, 
we find the calculated O~{\scriptsize{I}} and 
C~{\scriptsize{I}}  emission intensities are close to the
{Alice} observation by changing O$_2$ and CO relative 
abundances to 1 and 8\% with respect to water, 
respectively.  By varying the gas production rate and relative 
abundances of other species, we could not find a set of neutral parameters 
which can explain all the observed emission intensities 
simultaneously.

\begin{table*}
	\centering
	\begin{minipage}{16cm}	
		\caption{The comparison between model calculated 
			H  {\scriptsize{I}}, O  {\scriptsize{I}},
			and C  {\scriptsize{I}} emission intensities 
			and  {Alice} observations for different 
			input 
			conditions.}
		\resizebox{\columnwidth}{!}{%
			\begin{tabular}{lllllllllllllllllll}
				\hline 
				Date  & \ r$^c$ & H$_2$O & \multicolumn{2}{c}{H  
					{\scriptsize{I}} Lyman-$\beta$} & O$_2$$^b$   
				& \multicolumn{2}{c}{O  
				{\scriptsize{I}} 				
					1304 \AA } & \multicolumn{2}{c}{O  
					{\scriptsize{I}} 
					1356 \AA} & CO$_2$ $^b$ & CO$^b$ & 
				\multicolumn{2}{c}{C  {\scriptsize{I}} 1657 \AA} \\ 
				& (AU) &(s$^{-1}$ $\times$ 10$^{26}$)& 
				\multicolumn{2}{c}{(R)$^a$} & (\%)
				& \multicolumn{2}{c}{(R)}      
				&\multicolumn{2}{c}{(R)} & (\%) & (\%)   & 
				\multicolumn{2}{c}{(R)}  \\ 
				\hline 
				&   & &   Obs. & Cal.  &  & Obs. & 
				Cal. & Obs. & Cal. & & & Obs. & 
				Cal.	\\[-1pt]
				2015 March 29 & 1.99 & 2.0  & 12.3 $\pm$ 
				2.1   &  11.90 & 
				1.5 
				&  9.18 $\pm$ 1.96  & 1.1 & 7.24 $\pm$ 
				1.63   & 1.66 & 
				4 & 2 & 3.11  $\pm$ 1.93  &   0.08 \\
					
				& & 12.5  &    &  70.19 & 1.5 &   & 6.46 &   & 
				9.82 & 4
				& 2 &   &   0.48 \\  
				& & 12.5     &    & 69.90   & 1.0 &   & 5.16 &   
				& 7.36 & 4
				& 8 &   &   1.69 \\
				\hline
		\end{tabular}}
		\label{tab:results} 
		\footnotesize{$^a$The calculated emission 
			intensities are in Rayleigh};
		\footnotesize{$^b$Relative abundances with respect 
			to H$_2$O production rate}; 
		\footnotesize{$^c$Heliocentric Distance of the 
			comet}; Obs.~is Alice observed {\citep[see Table 2 
			of][]{Feldman18}} and Cal.~is model 
			calculated 
			emission intensities;
	\end{minipage}
\end{table*}

\subsection{Effect of input parameters on the calculated line 
emission intensities} 
\label{sec:effect-inputs}
\subsubsection{Neutral abundances}
\label{discus:neutrals}
ROSINA's DFMS and RTOF sensors observed that the CO and 
CO$_2$\ relative abundances  are 
dynamically varying in the coma due to inhomogeneous out-gassing
of nucleus  on 2015 May 15
\citep[see][and references there in]{Hoang17}.
To assess the impact of model calculated atomic carbon 
intensities on neutral composition, we have increased CO$_2$\ relative 
abundance 
from 3 to 6\% by 	keeping CO at 2\% with respect to 
water. For this new composition, the contributions of  both 
CO$_2$ and CO  photodissociative excitations are 
nearly equal (50\%) to the total C~{\scriptsize{I}} 1561 
and 1657 \AA\ emission intensities.

For the calculations made on 2015 May 15, the impact 
of molecular oxygen on the O~{\scriptsize{I}} emission
intensities has been studied by decreasing O$_2$ 
relative abundance from 4 to 0.5\%. Even with  O$_2$ 
density reduced by a factor 8, the molecular oxygen largely 
($>$50\%) controls the O~{\scriptsize{I}} emission 
intensities; the contribution of O$_2$ photodissociative 
excitation is five times higher than that of H$_2$O for 
O~{\scriptsize{I}} 1152 \AA. In this case, about 50--60\% of the 
total for O~{\scriptsize{I}} 1304 and 1356 \AA\ 
intensities are controlled by photodissociation of O$_2$. By increasing 
molecular O$_2$\ abundance 
higher
than 4\% with respect to water, the modelled  
O~{\scriptsize{I}} emission intensities are found to 
increase linearly and the contribution from 
H$_2$O photodissociation  can be completely neglected. 
In all these cases, electron impact 
dissociative excitation processes  { are negligible}  ($<$1\%)  in 
determining O~{\scriptsize{I}} emission 
intensities.   

\subsubsection{Cross sections}
\label{discus:xs}
\cite{Hans15} reported H$_2$O photodissociative excitation 
cross section producing H~{\scriptsize{I}} Lyman emissions 
in the wavelength region 375--825 \AA\ whereas, \cite{Wu79} 
measurement was limited to 600--800 \AA. By using the 
\cite{Wu79} measured cross section in the model, we find that 
the calculated photodissociative volume excitation rate of H$_2$O
is decreased by a factor 2.

\cite{Shirai01} recommended electron impact cross section 
of H$_2$O, producing H~{\scriptsize{I}} 
Lyman-$\alpha$ emission,  is higher than that of \cite{Itikawah2o}.
The cross section ratio of \cite{Shirai01} to 
\cite{Itikawah2o} at 100 eV (at 25 eV) is about 1.5 (5). 
By replacing \cite{Shirai01} recommended cross section with 
\cite{Itikawah2o}, no significant change in the model 
calculated total H~{\scriptsize{I}} emission  intensities 
is seen since most ($>$95\%) of this emission is controlled 
by solar resonance fluorescence of hydrogen atoms.

Based on \cite{Wu88} experimentally determined 
yield, we have assumed the cross sections for 
photodissociative excitation of H$_2$O\ producing   
O~{\scriptsize{I}} 1304 and 1356 \AA\ emissions are same as 
that of  
O~{\scriptsize{I}} 1152 \AA.  As shown in 
Figures~\ref{fig:emis-inten-oa} to \ref{fig:emis-inten-oc}, 
dissociative excitation  H$_2$O  is the second 
largest process which contributes to the total  O~{\scriptsize{I}} 
emission intensity. Since most of the UV photons 
in the wavelength region 600--800 \AA\  ionize H$_2$O\ rather than
producing emissions via dissociative excitation 
 \citep{Huebner92}, we do not expect the actual cross section value 
 to be an order of 
magnitude higher to the assumed value in this model. We found 
that the contribution of H$_2$O\ to total  
O~{\scriptsize{I}} emissions is not significant ($<$1\%)
even on increasing photodissociative  cross sections by a factor 2.

The dissociative excitation of O-bearing species, which leads to O($^5$P)
formation, produce O~{\scriptsize{I}}  7774 \AA\ emission 
subsequently contribute to  
O~{\scriptsize{I}} 1356 
\AA\ line. We have considered the  O($^5$P) formation channel 
for the photodissociation of O$_2$\ and also for the 
electron impact excitation of H$_2$O in the model. The 
additional 
contribution of O~{\scriptsize{I}} 7774 \AA\ to 
O~{\scriptsize{I}} 1356 \AA\ emission intensity is about 10\% via 
photodissociation of O$_2$ and 50\% for electron impact on 
H$_2$O. We could not include H$_2$O photodissociation which 
produce O~{\scriptsize{I}} 7774 \AA\ emission in the model 
due to non-availability of cross section. We do not expect 
this dissociative excitation mechanism would significantly 
alter our calculated O~{\scriptsize{I}} 1356 \AA\ intensity 
since the photodissociation of O$_2$ majorly ($>$90\%) controls 
this emission.

\cite{Zhou14} measured O$_2$ dissociative excitation cross 
section producing O($^3$S) for a limited   
wavelength region (815 -- 850 \AA). This cross section  is
smaller by an order of magnitude compared to the 
measurement of \cite{Carlson74}. 
By using \cite{Zhou14} cross section,  we find that
photodissociative excitation rate of O$_2$ producing  
O~{\scriptsize{I}} 
1304 \AA\ emission  is decreased by two orders of magnitude.
In this case photodissociation of H$_2$O\ is a dominant 
production source in the formation of  O~{\scriptsize{I}} 
1304 emission.
However, \cite{Zhou14} measured photodissociative 
excitation 
cross section restrained to a small wavelength band which 
does not implicate the total photochemical process in 
cometary coma. 

We have not accounted for CO$_2$ and CO 
photodissociative excitation cross sections, which produce 
O~{\scriptsize{I}} 1152 and 1356 \AA\ emissions,  in our 
model  due to the lack of reported cross sections in the 
literature. On assuming these emissions cross sections are 
same as that of 1304 \AA\ \citep[as measured by][]{Wu79a}, 
we could not find any significant change in the total 
O~{\scriptsize{I}} 1152 and 1356 \AA\ emission intensities. 
This is mainly due to photodissociation of O$_2$ is the 
dominant production source of O~{\scriptsize{I}} 1304 \AA\ 
line, which is an order of magnitude higher compared to any 
other dissociative excitation processes (see 
Figure~\ref{fig:emis-inten-ob}).

\begin{figure}[htb]
	\centering	
	\includegraphics[width=\columnwidth]{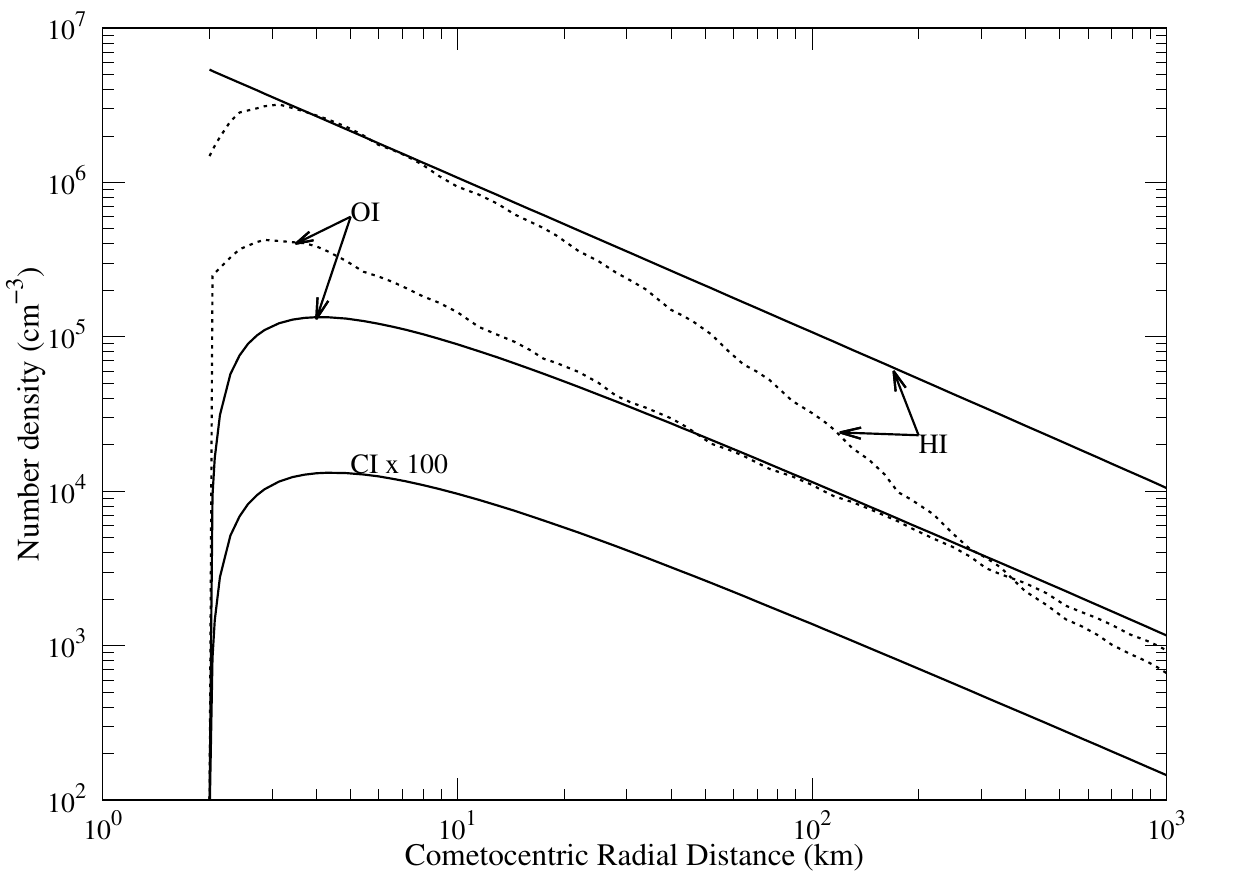}
	\caption{The calculated  number density 
	profiles (solid lines) of atomic hydrogen, oxygen, and carbon in 
	comet 67P/C-G for water production rate of 5 $\times$ 
	10$^{27}$ s$^{-1}$ and for 3\% CO$_2$, 2\% of CO and 
	4\% of O$_2$ when comet was at 1.56 AU. The dashed 
	lines represent the DSMC modelled H~{\scriptsize{I}} 
	and O~{\scriptsize{I}} number density profiles from \cite{Tenishev08}. 
	Atomic 
	carbon number densities are plotted after multiplying 
	by a factor 100.}
\label{fig:minor_den}
\end{figure}

\subsubsection{Minor atomic densities}
\label{results:mden}
 {The model calculated atomic carbon, oxygen, and hydrogen number 
 density profiles are plotted in 
Figure~\ref{fig:minor_den}, for a water production rate of 
5 $\times$ 10$^{27}$ s$^{-1}$, along with  DSMC modelled 
H~{\scriptsize{I}} and O~{\scriptsize{I}} number densities  
\citep{Tenishev08}. These atomic oxygen 
(hydrogen) densities are consistent with DSMC values for 
radial distances above 10 km (below 20 km). 
The DSMC model 
accounts for momentum exchange between water and highly 
energetic hydrogen atoms produced via dissociation while calculating 
H~{\scriptsize{I}} densities, whereas 
our model number densities are based on  
Haser's radial expansion model does not account for 
collisions, which is the main reason for the discrepancy 
between number density profiles. Even on using DSMC 
modelled atomic hydrogen number density profile in the 
model, we find solar resonance fluorescence is the dominant source 
of H~{\scriptsize{I}} Lyman-$\alpha$ emission, whereas, 
photodissociation of H$_2$O is the significant excitation 
source of H~{\scriptsize{I}} Lyman-$\beta$ and $\gamma$ 
emissions for radial distances below 1000 km.}

\section{Discussion}
\label{sec:discussion}
The primary assumption in the analysis of  {Alice} 
observed H~{\scriptsize{I}}, O~{\scriptsize{I}}, and 
C~{\scriptsize{I}}  emissions on comet 67P/C-G\ is that the 
electron impact excitation is the major source of these 
emissions \citep{Feldman15, Feldman16, Chaufray17, 
Feldman18}. However, earlier modelling works have shown 
that these emissions are majorly governed by resonance 
fluorescence of solar photons at corresponding wavelengths 
and dissociative excitation of cometary volatiles by photon 
and photoelectron \citep{Bhardwaj96, Combi00, Combi04, 
Feldman04}. Our photochemical model calculations,
 when the comet was at 1.56 AU from the Sun,
 show that 
photodissociative excitation is the major source of these 
emissions (except for H~{\scriptsize{I}} Lyman-$\alpha$) 
for radial distances below 100 km and the role of electron 
impact reactions is small 
(see~Figures~\ref{fig:hydro-emis}--\ref{fig:carb-emis-a}).

\subsection{On the role of suprathermal electrons in producing 
atomic emissions}
Our modelled suprathermal electron fluxes on 2015  
June 5 and July 26, when comet 67P/C-G was at 1.5 and 
1.26 AU from the Sun with respective gas production rates 
as 3 $\times$ 10$^{27}$ and 1 $\times$ 10$^{28}$ s$^{-1}$,
are consistent with RPC-IES 
daily-averaged measurements in the energy range 10--60 eV 
(see Figure~\ref{fig:cmp_eflx}). But the modelled flux on 
2015 March 23, when the comet was at 2 AU from the Sun with 
a gas production rate of 2 $\times$ 10$^{26}$ 
s$^{-1}$, is smaller compared 
to RPC-IES measurement for the energy above 20 eV 
(by a factor 5 or more). This 
discrepancy could be mainly due to less frequent collisions 
between suprathermal electrons and cometary species in a low 
gas production rate comet. Since our model calculations are 
based on the local degradation approximation, the 
calculated suprathermal electron flux at 2 AU heliocentric 
distance is not in agreement  with the RPC-IES observation.

The threshold energies for dissociative excitation 
of major cometary species, which produce atomic 
emissions, is more than 15 eV (see Fig.~\ref{fig:ele-csc}).
Hence, the lower ($<$ 15 eV) component of suprathermal electron flux 
spectrum does not play any role in determining the volume 
emission rates of atomic 
excited states. Even though the electron impact excitation 
cross section is maximum around 100 eV, since the 
suprathermal electron flux for energies more than 100 eV 
decreases rapidly by several orders of magnitude 
\citep{Madanian16,  Madanian17, Clark15, Broiles16} the 
role of this 
 high ($>$100 eV) energetic component in producing atomic 
 emissions can be neglected. If electron impact is the only 
 excitation mechanism in the coma, then about 80\% 
 the atomic emission intensities is determined by 
 the suprathermal electron flux in the energy range 
 15--70 eV.

We have also quantified 
the contribution of solar photons and suprathermal 
electrons in producing atomic emissions for 2015 March 
29 observation,  when the
comet was at 2 AU and having a gas production rate 
of 2 $\times$ 10$^{26}$ s$^{-1}$ with CO$_2$, CO 
and O$_2$ volume mixing ratios 4, 2, and 1.5\%, respectively,
by varying  {the RPC-IES measured suprathermal electron flux magnitude 
as inverse distance to nucleus}  in the cometary coma.
We find the role of solar photon is higher by a factor of 5
(2.5)
or more than electron impact excitation 
in producing atomic H~{\scriptsize{I}} and 
O~{\scriptsize{I}} (C~{\scriptsize{I}}) emissions. 
Similarly, on using  RPC-IES measured suprathermal electron flux 
in the model at 1.5 AU, we find no significant change in the
calculated emission intensities.
Based on these calculations, we suggest that the role of 
photons is considerably higher 
than suprathermal electrons (about a factor 2 or more) 
in governing atomic emission 
lines when comet 67P/C-G was at heliocentric distances 2 
AU or below.

\subsection{Implications for formation processes of 
H~{\scriptsize{I}}, O~{\scriptsize{I}}, and 
C~{\scriptsize{I}} emissions}
Our calculated volume emission rates  
clearly show that photodissociation of H$_2$O contributes 
to the total H~{\scriptsize{I}} 
Lyman-$\beta$ emission by an order of magnitude 
higher compared to 
solar resonance fluorescence close to the cometary surface (see 
Figure~\ref{fig:hydro-emis}).
\cite{Feldman18} also found that the contribution 
from resonance scattering for H~{\scriptsize{I}} 
Lyman-$\beta$ emission is negligible in the Alice 
observations. 
It should be noted that the Rosetta 
was orbiting at radial distances around 90 km on 2015 
March 29 while Alice was making nadir observations, hence 
the contribution from solar resonance fluorescence in the 
nadir observed HI Lyman-$\beta$ intensity can be neglected.

Even if we neglect the electron impact 
dissociative excitation completely, our
model calculated H~{\scriptsize{I}} Lyman-$\beta$ line 
intensity can be in agreement with  the  {Alice} 
observations (see 
Figures~\ref{fig:hydro-emis} and \ref{fig:emis-inten}). 
This clearly suggests that the photodissociative excitation of 
H$_2$O is the major source for the Alice
observed H {\scriptsize{I}} Lyman-$\beta$ emission below
 50 km projected distances and the contributions 
from electron impact excitation and solar
resonance fluorescence can be neglected. As discussed 
in Section~\ref{sec:results},  the electron impact 
on H$_2$O could not be a dominant excitation process in producing 
H~{\scriptsize{I}} emissions even on using higher 
cross sections (by a factor of 2 to 4 larger compared 
to \cite{Itikawah2o} recommended values). Since photodissociative 
excitation of H$_2$O is the major production mechanism 
for the observed H  {\scriptsize{I}} Lyman emissions, we suggest that the 
observed emission intensities are suitable to
derive water production rates rather than the electron density 
when the comet is closer to the perihelion.

The model calculations show that despite having 
H$_2$O as the dominant O-bearing species in cometary 
coma, photodissociation of O$_2$ is the major source for
producing  O {\scriptsize{I}} emissions viz.,
1152, 1304, and 1356 \AA\ (see 
Figures~\ref{fig:emis-inten-oa}, \ref{fig:emis-inten-ob}, 
and \ref{fig:emis-inten-oc}). This is mainly due to the large  
difference (more than three orders of magnitude, see Figure 
\ref{fig:ele-csc}) between H$_2$O and 
 {O$_2$} photodissociation cross sections. 
Moreover, the difference in the magnitudes of solar flux in 
 dissociative wavelength regions 150--650 \AA\ (for H$_2$O)  
and 700--850 \AA\ (for O$_2$) also play an important role 
in determining the photodissociation rates and subsequently 
O {\scriptsize{I}}  emission intensities.

The photochemical origin of O {\scriptsize{I}} 1152 \AA\ 
emission in cometary spectra is  thought to be  due 
to solar resonance fluorescence of O($^1$D -- $^0$D), which 
is  an analogous  emission mechanism of C($^1$D -- 
$^1$P$^0$) at 1931 \AA\ \citep{Bhardwaj99a,Combi04}. This emission
line has been detected on comet C/2001 A2 (LINEAR) by Far 
ultraviolet spectroscopic explorer (FUSE) satellite. 
The observed emission intensity could not be explained by 
the previously suggested excitation process and is attributed 
to the presence of an enigmatic source 
\citep{Feldman02,Combi04}. Our model calculations 
show that photodissociation of O$_2$ and H$_2$O  can 
populate oxygen atoms directly in $^0$D state (see 
Figure~\ref{fig:emis-inten-oa}). Based on our previous 
O($^1$D) model calculations done for 67P/C-G\ and several 
other comets \citep{Raghuram16,Bhardwaj12,Raghuram13}, we 
argue that the resonance fluorescence is  not a significant 
emission process for O {\scriptsize{I}} 1152 \AA\ 
line in comets. Moreover, the radiative lifetime of 
O($^1$D) is about 110 s, unlike C($^1$D) state \cite[which 
is 4077 s,][]{Tozzi98}, during which it has to fluoresce 
solar photons at wavelength 1152 \AA. Our model 
calculations clearly indicate that even on reducing 
O$_2$ abundance to 0.5\%, the total
O~{\scriptsize{I}} 1152 \AA\ emission intensity is majorly
controlled by the photodissociation of O$_2$ ($>$90\%). 
Hence, we suggest that this emission can be considered 
as a proxy to measure the O$_2$ abundance in 67P/C-G coma.

Among the atomic oxygen line emissions considered 
in this work, the O~{\scriptsize{I}} 1356 \AA\ is due to a 
spin forbidden  transition ($^5$S $\rightarrow$ $^3$P). 
Owing to a very small  g-factor ($<$10$^{-8}$ 
s$^{-1}$), the observation of this cometary line 
is mainly attributed to electron impact dissociative 
excitation processes and has been used to establish the 
presence of photoelectrons in the cometary coma 
\citep{Feldman04}. Our model calculations (see 
Figure~\ref{fig:emis-inten-oc}) show that photodissociation 
of O$_2$ is the major source of this emission compared 
to earlier assumed electron impact excitation of O-bearing 
species.  {If our assumed emission cross section for  
the photodissociation of H$_2$O is realistic (see 
Section~\ref{sec:model-inputs}), 
then} in the case of comets 
deprived of molecular 
oxygen, dissociative excitation of H$_2$O\ can be a major 
production mechanism in determining the O {\scriptsize{I}} 
1356 \AA\ emission intensity. Our model calculations 
clearly show that electron impact excitation on 
O-bearing species plays a minor role ($<$1\% to the total)  in 
producing this emission. Hence,  we suggest that the 
observation of O {\scriptsize{I}} 1356 \AA\ line 
necessarily need not represent the presence of electron 
impact excitation reactions in the cometary coma.

 {To our knowledge, there are no reported cross sections for OI 1356 
 \AA\ emission for photodissociation of H$_2$O in the literature. The 
 small value of 
cross section \citep[$<$5 $\times$ 10$^{-21}$ cm$^2$,][]{Wu88} and long 
radiative lifetime of the excited state ($\sim$180 ms)  inhibits the 
determination of dissociative excitation cross section of H$_2$O over wide 
range of wavelength in the laboratory \citep[][]{Makarov03,Kanik03}.}
 Moreover, the formation thresholds of O($^5$S) and O($^3$S) (which 
are the corresponding excited states of  O~{\scriptsize{I}} 1356 and 1304 \AA\ 
emissions) are about 14.2 eV (873 \AA) and 14.5 eV (854 \AA), 
respectively, and higher than the ionization threshold of H$_2$O (12.6 eV).
\cite{Haddad86} have determined that the H$_2$O photoionization yield for the 
wavelengths smaller than 600 \AA\ (20.7 eV) is almost equal to one. Hence, the 
absorption of photons by H$_2$O in the energy range 14.5--20.7 eV majorly leads 
to the formation of O~{\scriptsize{I}} emission via dissociative excitation.  
 {Due to the significant difference in the photodissociation cross 
sections (about an order of magnitude), O$_2$ can be a more efficient source of 
O~{\scriptsize{I}} emissions compared to H$_2$O.}

 {By using a gas production rate of 1.3 $\times$ 10$^{30}$ s$^{-1}$ and 
for the fractional composition of H$_2$O, CO and CO$_2$ as
 80\%, 20\%, and 3\%, respectively,} the model calculations on comet 1P/Halley 
 by \cite{Bhardwaj96}  showed that 35\% of  O~{\scriptsize{I}} 
1304 \AA\ emission intensity  {in the inner coma} is governed by electron 
impact on atomic oxygen  and remaining is from resonance 
fluorescence. However, this model does not account for photodissociative 
excitation of O-bearing species producing OI 1304 \AA\ emission. But due to the 
presence of significant amount of O$_2$ ($\sim$4\% relative to H$_2$O) in 
67P/C-G,  photodissociation of molecular oxygen can be a more effective 
mechanism in producing atomic oxygen compared to any other formation 
processes as discussed by \cite{Bhardwaj96}. We have 
evaluated the contribution of atomic oxygen resonance 
fluorescence producing 1304 \AA\ emission and found to be 
$\sim$1\% to the O$_2$ photodissociation.  Our model 
calculated volume emission rates show that 
photodissociation of O$_2$ is the major source
for O~{\scriptsize{I}} 1304 \AA\   and the role of 
dissociative excitation of other oxygen-bearing species is small
(by more than an order of magnitude, 
see Figure~\ref{fig:emis-inten-ob}). 
As discussed in Section~\ref{sec:effect-inputs}, even on 
reducing O$_2$\ relative abundances to 0.5\% with respect 
to water, photodissociation of O$_2$\ is the dominant 
($>$50\%) source of O {\scriptsize{I}} lines. The 
observations of \cite{Bieler15} showed that molecular 
oxygen abundance can be even 15\% with respect to water, in 
this case, the contribution from all other sources can be 
completely neglected. Based on our model calculations, we 
suggest that atomic oxygen line emissions are directly 
linked to O$_2$ molecular abundance and hence can be 
used to probe its sublimation rate from the nucleus.

In this work we have considered CO and CO$_2$ are the 
only carbon bearing species as the source of C 
{\scriptsize{I}} emissions. The model calculations
show that the photodissociation 
rate of CO is higher by an order of magnitude  compared to  
that of CO$_2$ producing  C  {\scriptsize{I}} emissions  (see 
Figure~\ref{fig:carb-emis-a}). This is because the 
dissociation of CO 
leads to direct formation of excited atomic carbon which is 
a slow process in the case of CO$_2$ dissociative excitation. 
Using photon (electron) cross sections, we evaluated that 
the atomic carbon in ground state 
produces via CO dissociation with a rate about 400 (2000) 
times higher than that of CO$_2$.
These calculated ratios suggest that CO is a potential source of 
atomic carbon in the cometary coma than CO$_2$ either 
by photon or electron impact dissociation. 
 {\cite{Hassig15} measurements show that CO/CO$_2$ 
ratio can be as high as 5 due to heterogeneity of the 67P/C-G cometary 
coma in which case the contribution from CO$_2$ can be 
completely neglected in the formation of C~{\scriptsize{I}} 
emission intensities. The model calculated emission 
intensity ratios of C {\scriptsize{I}} 1657/1561 \AA\
for the photodissociation of CO and CO$_2$ are 2 and 1.7, respectively (see 
Figure~\ref{fig:carb-emis-a})}. The resonance fluorescence 
efficiencies ratio of atomic carbon for  C~{\scriptsize{I}} 
1657/1561 lines is about 2.8  \citep{Woods86}. These
calculated ratios indicate that if CO and/or CO$_2$ are the 
primary sources then the C~{\scriptsize{I}}  1657 
\AA\ emission should always be more intense than that 
of C~{\scriptsize{I}} 1561  \AA, which has actually been 
observed in several  {Alice} spectra 
\citep{Feldman15,Feldman18,Noonan18}. This show 
that C~{\scriptsize{I}} 1657 \AA\ emission
line is largely controlled either directly (via 
dissociative excitation) or indirectly (via dissociation 
leads to atomic carbon which further resonates solar 
photons at wavelength 1657 \AA) by CO rather than CO$_2$.  
 {Our calculations in Table~\ref{tab:results} also show that
when Q$_{H_2O}$ =  12.5 $\times$ 10$^{26}$ s$^{-1}$ and for 4\% CO$_2$ relative 
abundance, there is  a linear increment 
in the modelled C~{\scriptsize{I}} 1657 \AA\ emission intensity with increasing 
the CO relative abundance from 2 to 8\%. In this case, for a 
fixed CO abundance (4\% relative to water) and by varying the CO$_2$ abundance 
by a factor 5, we find the impact of CO$_2$ on the 
the C~{\scriptsize{I}} emission intensities is very small ($<$5\%). This 
calculation suggests that CO can a significant source of 
C~{\scriptsize{I}}  emissions than CO$_2$.}
Hence, we suggest that the observed C~{\scriptsize{I}}  
emission intensities  {are more directly linked with photodissociation of 
CO rather than that of CO$_2$ and} can be used to constrain its abundance 
in the cometary coma.

For determining the  H~{\scriptsize{I}}, O~{\scriptsize{I}}, and 
C~{\scriptsize{I}} nadir emission intensities on 2015 March 29, we have 
considered 
the variability in chemical composition of cometary coma
as seen by ROSINA and the suprathermal electron flux as measured RPC-IES 
in the model.
We find our calculated H~{\scriptsize{I}}  Ly-$\beta$ emission intensity 
on 2015 March 29 is in agreement with the Alice observed 
values when we use the lower limit of ROSINA measured water 
production rate, whereas the modelled O~{\scriptsize{I}} 
emissions are not consistent (see Table~\ref{tab:results}).
But when we varied the gas production rates and mixing ratios 
to the upper limits of measurements, 
our model calculated O~{\scriptsize{I}}   
emission intensities are close to the observations.
The discrepancies between modelled and  observation 
nadir intensities on 2015 March 29 could be due the 
strong chemical heterogeneous composition of coma 
along the observed line of sight of  Alice spectrometer.
It should be noted that we have used Rosina measured gas 
production rates and relative mixing ratios which are 
derived at spacecraft position. But along the Alice line of 
sight the volume mixing ratios of different species may be 
varying due to chemical inhomogeneity in the coma as 
discussed by \cite{Keeney17}.

\subsection{The effect of electron impact excitation at large ($\sim$3 AU) 
heliocentric distances on the emission 	intensities}
 {RPC-IES observations at large ($\sim$3 AU) 
 	heliocentric distance  have shown that
suprathermal electron flux was dynamically varying in the 
cometary coma  \citep{Clark15, Broiles16, Madanian16, 
Madanian17}.} The initial plasma studies of 
\cite{Galand16}, when the comet was at 3 AU from the Sun, 
have shown that electron impact ionization rate is higher 
than photoionization rate by a factor 2 to 10. Using 
RPC-IES measured suprathermal electron flux and H$_2$O 
electron impact ionization cross section, \cite{Galand16} 
have also shown that electron impact H$_2$O ionization  
frequency significantly  varies (between 0.1--8 $\times$  10$^{-7}$ 
s$^{-1}$)  within a few hours of observation. This is 
mainly due to the low sublimation rate (10$^{26}$ s$^{-1}$) 
of the comet that leads a highly rarefied medium around the 
nucleus and the hot component of photoelectrons could not 
exchange energy to the surrounding neutral species in the 
cometary coma.  However, the plasma studies were done by 
\cite{Heritier17, Heritier18} on 67P/C-G, when the comet 
was around at perihelion distance  {(1.24 AU)} with a gas production 
rate of about  10$^{28}$ s$^{-1}$, show that electron 
impact ionization is non-significant compared to 
photoionization. This study indicates that in a 
well-developed cometary coma of a comet, photon impact 
reactions play a significant role compared to electron 
initiated processes. {Similarly, as the comet was moving towards the 
Sun, the wide Angle camera of Optical, 
Spectroscopic, and Infrared Remote Imaging System (OSIRIS) and Alice onboard 
Rosetta instruments also observed the changes in the intensities of various 
spectral emissions of cometary species \citep{Bodewits16, 
Feldman18}. By analysing the Alice observed far-ultraviolet spectra, 
\cite{Feldman18} have studied the evolution of cometary coma and suggested 
that 
various emission features of 67P/C-G are majorly driven by electron impact 
dissociative excitation processes. When the comet was moving towards 
the Sun from $\sim$2 to 1.5 AU, \cite{Bodewits16} have noticed that 
the OSIRIS observed emission intensities of various species are 
decreased though the gas 
production rate was increasing in this period. These observations 
also suggest that  the collisions between 
electrons and neutrals is increased significantly with 
the increase in local neutral density which leads to a
strong electron cooling and  a transition of inner coma from electron 
to photon driven excitation.}

It should be noted that the
electron impact ionization cross section is more 
than  an order of magnitude higher compared to dissociative 
excitation of H$_2$O\ producing H~{\scriptsize{I}} Lyman 
emissions \citep{Itikawah2o,Makarov04}. Hence, 
the high energetic photoelectron mainly turns water 
molecule into ion rather than to produce atomic emissions 
via dissociative excitation. To assess the 
contribution of electron impact excitation of H$_2$O 
producing H~{\scriptsize{I}} emissions at heliocentric 
distance 3 AU, we have scaled \cite{Galand16} determined 
ionization rates with the mean electron impact cross 
section ratio of H~{\scriptsize{I}} emission to H$_2$O 
ionization. Our determined electron  impact volume emission 
rates producing H~{\scriptsize{I}} lines are higher by 
one to three orders of magnitude compared to modelled 
H$_2$O photodissociative excitation rates at this 
heliocentric distance. This calculation suggests that 
electron impact excitation mechanisms are largely driving 
the emission intensities compared to photodissociative 
processes in comet  {67P/C-G}  at large ($\sim$3 AU) 
heliocentric distances. In this case, the observed 
emission intensity can be used to derive the electron 
density in the cometary coma. But  when  {the comet 
67P/C-G} reaches closer to the Sun, the magnitude of hot 
component suprathermal electron flux reduces due to strong 
collisional degradation and the photodissociative excitation reactions 
play a major role in determining the atomic emission 
intensities. 

\subsection{On the role of chemical heterogeneity of cometary coma}
 {Several observations have shown that the distribution of neutral 
species 
in the cometary coma is heterogeneous due to uneven distribution of source 
regions on the nucleus \citep{Hassig15, Hoang17, Fougere16}. 
The 
non-uniformity in the neutral distribution along the Alice line of sight  can 
also impact the observed emission intensities. During the Alice observation on 
25 May 2015, the modelled column density of H$_2$O (6 $\times$ 10$^{15}$ 
cm$^{-2}$) is consistent with the VIRTIS-H observation (2--7 $\times$ 
10$^{15}$ cm$^{-2}$)  \citep[see Figure 4 of ][]{Chaufray17}. Considering the 
variability of H$_2$O column density as observed by VIRTIS, we evaluated the 
effect of 
neutral inhomogeneity on the modelled emission intensity. By decreasing the 
H$_2$O production rate by a factor of 3, we find 
the modelled H~{\scriptsize{I}} emission intensities are decreased by a factor 
of 2. In this case, solar photons are the major driving source of 
H~{\scriptsize{I}} emissions and the role of electron impact is negligible.} 

 {The mixing ratios of other O-bearing species with respect to H$_2$O are 
dynamic
around the nucleus and also play an important role in determining the 
emission intensities. The earlier observations of \cite{Hassig15}, when the 
comet was beyond 3 AU, have shown that CO$_2$/H$_2$O density ratio varied by an 
order of magnitude in the coma. But as the comet passed the inbound equinox 
(2015 May 10), the CO$_2$/H$_2$O and the CO/H$_2$O  
density ratios around the nucleus, due to the spatial heterogeneity, are less 
than 7\% and 15\%, 
respectively, \citep[see Figure 2 of][]{Hoang19}.  By varying the CO and 
CO$_2$ 
abundances 
within the measured limits of RTOF/Rosina, we find the contribution CO is 
more significant ($>$50\%) than CO$_2$ in determining the C~{\scriptsize{I}} 
emission intensities. However, as the observations of \cite{Hoang19} show that 
the 
dynamical variability of these species is significant (relative abundances are 
$\sim$1 with respect to 
H$_2$O) when comet approach the outbound equinox, and  the observed emission 
intensity 
can significantly be altered by the neutral composition. Hence, the effect of 
chemical heterogeneity of the cometary coma should be considered while deriving 
the gas production rate based on the observed emission intensity at different 
heliocentric distances. Calculation of atomic emission intensities using a 
realistic 
cometary neutral atmospheric model, which accounts for the non-homogeneous 
distribution and dynamical variation of neutral species, is outside the scope 
of this paper.}

\subsection{On H~{\scriptsize{I}} Lyman-$\alpha$/ 
Lyman-$\beta$ emission intensity ratio}  
 {The initial Alice observations made on comet 67C-G,  {at heliocentric 
 distances beyond 3 AU 
 pre-perihelion,} have shown that the H~{\scriptsize{I}} 
Lyman-$\alpha$/Lyman-$\beta$ emission intensity ratio is 
around 5 \citep{Feldman15} which is close to the 
\cite{Makarov04} determined electron impact emission cross 
section ratio 7 at 100 eV. Hence, this observed emission 
ratio has been used as a confirmation that electron impact 
excitation reactions are the dominant source of atomic 
emissions. Our { model calculations, when the comet was at 1.56 AU,} 
shows 
that solar resonance 
fluorescence is the major source of H~{\scriptsize{I}} 	
Lyman-$\alpha$ 
emission than  H$_2$O dissociative excitation by photons 
and electrons (See Fig.~\ref{fig:hydro-emis}). The solar
resonance fluorescence of H~{\scriptsize{I}} Lyman-$\alpha$
mainly depends 
on the atomic hydrogen density which varies as a function 
of water production rate and also due to kinetic effects in 
the cometary coma as described by \cite{Tenishev08}.
By developing a global kinetic model developed for 
comet 67P/C-G, \cite{Tenishev08} have shown that the 
diurnal variability in the gas 	production rate can lead to 
a significant momentum 	exchange between energetic hydrogen 
atoms and water molecules near nucleus which results in 
change (by an 	order of magnitude) in atomic hydrogen 
density within 100 	km radial distances. . This indicates that the 
observed H~{\scriptsize{I}} Ly-$\alpha$/Ly-$\beta$ 
emission ratio can also change as  the number density of 
atomic hydrogen varies in the inner coma.  We have used 
Haser fitted atomic hydrogen number density profile in the 
model to evaluated H~{\scriptsize{I}} 
Ly-$\alpha$/Ly-$\beta$ emission ratio.
 {The modelled H~{\scriptsize{I}} Ly-$\alpha$/Ly-$\beta$ ratio value 
on 25 May 2015 is greater than 100 near nucleus and increases with 
cometocentric projected distance (see 
Figure~\ref{fig:emis-inten}). If resonance fluorescence is the only excitation 
mechanism in the coma, then the observed  H~{\scriptsize{I}} 
Ly-$\alpha$/Ly-$\beta$ emission ratio should be around 750 
\citep{Gladstone10}. Since dissociative excitation reactions of H$_2$O are 
playing 
a significant role close to the nucleus, our modelled emission ratio is smaller 
than the 
resonance fluorescence emission ratio.} 
 As explained in 
Section~\ref{results:mden}, even on using 
DSMC 
modelled atomic hydrogen number densities we find solar 
resonance fluorescence as the significant source of  
H~{\scriptsize{I}} Lyman-$\alpha$ emission. So we suggest 
that the observed 
H~{\scriptsize{I}}  Ly-$\alpha$/Ly-$\beta$ emission ratio 
not necessarily  confirms that the electron impact reactions 
are significant in producing atomic emissions. The 
calculation of atomic hydrogen number density by 
accounting for  kinetic effects of the cometary coma is 
beyond the scope of present work.}

\subsection{Atomic oxygen line emission intensity ratio}
Several factors play a role in the Alice 
observed O~{\scriptsize{I}} emission intensity ratios along 
its line of sight. Alice observed 
O~{\scriptsize{I}} 1304/1356 emission intensity ratios on 
2014 Nov. 29, 2015 Jan. 30, and 2015 March 29 as 2.91,  
0.67, and 1.27, respectively \citep{Feldman18}. During the 
cometary outburst period i.e., on 2015 May 22--24, 
the  observed that this emission ratio is varying 
between 1 and 2 \citep{Feldman16}. It is difficult to 
explain this variability in the observed O~{\scriptsize{I}} 
emission ratio by invoking dissociative excitation of 
a single O-bearing species.  If the cometary coma is 
homogeneously mixed with molecular oxygen and 
photodissociation of O$_2$ is the major production source 
of these emissions, then our calculations suggest that 
O~{\scriptsize{I}} 1304/1356 ratio should be around 0.5.  
Unfortunately, there are no measured  H$_2$O 
photodissociative  cross sections to determine 
O~{\scriptsize{I}} 1304/1356 \AA\ brightness ratio. If our 
assumed photodissociation cross sections of H$_2$O, {which are estimated 
based on the experimental yield of \cite{Wu88},} are  {realistic}
and photodissociation of H$_2$O play significant role in 
producing these emissions, then O~{\scriptsize{I}} 1304/1356
emission intensity should be around 1. These ratios  
suggest that  H$_2$O and O$_2$ photodissociative excitation 
reactions alone can not produce the observed higher ($>$1) 
emission ratios.

{If  electron impact is the only dominant 
excitation source in the cometary coma, as observed on comet 67P/C-G at 
larger heliocentric distances ($>$3 AU), then the measured 
H$_2$O (O$_2$) electron impact cross section ratios suggest 
that the O~{\scriptsize{I}} 1304/1152  and  1304/1356 
 emission intensity ratios should be always $\sim$1.21 
 (0.07)  and  $\sim$2.5 (0.5), respectively  
 \citep{Makarov04,Kanik03}.}  {If CO$_2$ (CO) is the dominated species 
 in the coma, then for electron impact as dominant excitation mechanism the  
O~{\scriptsize{I}} 1304/1152  and  1304/1356 emission ratios should be around 
0.5 (2) and  2 (1.17), respectively \citep{Ajello71-co,Itikawaco2,Kanik95}.}    
Hence, based on the previously  mentioned theoretical ratios it is  difficult 
to explain  the observed variability in O~{\scriptsize{I}} 1304/1356 
emission ratios  {when} the cometary composition  
significantly changes  {with different O-bearing species along the Alice 
observed line of sight. Moreover, 
resonance fluorescence of atomic oxygen can also be an important excitation 
mechanism that can significantly alter the observed O~{\scriptsize{I}} 
1304/1356 
emission ratio. Hence, we suggest that O~{\scriptsize{I}} 1304/1356  emission 
ratio alone can not indicate the excitation source in the cometary coma}.

For the cometary coma, where the electron impact 
excitation is major excitation source, the molecular oxygen can be a potential 
source for  
 O~{\scriptsize{I}} emissions compared 
to H$_2$O due to large difference in electron impact 	
cross sections for electronic impact excitation. The electron impact cross 
section of 
O$_2$ producing O~{\scriptsize{I}} 1304 \AA\ (1356 \AA) 
emissions is higher by an (about two) order(s)  of 
magnitude compare to that of H$_2$O in the energy range 
20--40 eV where the peaks occur in the suprathermal 
electrons spectrum (see Figure \ref{fig:ele-csc}). Our 
calculated volume emission rates, for 4\% O$_2$ relative 
abundance, also suggest that the electron impact on O$_2$ 
and H$_2$O  {can be } equally potential sources in producing 
O~{\scriptsize{I}} 1304 and 1356 \AA\ emissions (See 
Figs.~\ref{fig:emis-inten-oa} and \ref{fig:emis-inten-oc}).
Hence, the presence of small  amount  of 
molecular oxygen  in cometary coma ($\sim$5\%)  can be a significant 
source for O~{\scriptsize{I}} emissions than 
H$_2$O.

The volume mixing ratios O$_2$/H$_2$O is one of the 
importance parameters which can determine the Alice 
observed line of sight O~{\scriptsize{I}}
emission ratios. As mentioned earlier, molecular 
oxygen has been detected in 67P/C-G with an average 
relative abundance of 4\% with respect to H$_2$O and found 
to vary   ($\sim$0.1--15\% relative to H$_2$O) 
in the coma during the Rosetta mission period 
\citep{Bieler15}. It should be noted 
that Rosina derived relative mixing ratios are based 
on the measurement of neutral gas densities at spacecraft
position. Using Alice spectrometer, \cite{Keeney17} found 
that the O$_2$/H$_2$O relative abundance ratio in the 
cometary coma is varying between 11--68\%, with a mean 
value of 25\%, by studying H$_2$O and O$_2$ absorption 
spectra of cometary coma against far-UV continuum of 
stellar radiation. Alice derived abundances depend on the 
column density of neutrals along the line of sight 
observation. Hence, the variability in O$_2$ and H$_2$O 
along the Alice line of sight can significantly affect the  observed 
atomic oxygen emission intensities and their ratios.

The observed  {limb }oxygen emission ratio also depends on 
the
viewing geometry of line of sight. This can be understood from our 
modelled O~{\scriptsize{I}} 1304/1356 and 1152/1304 \AA\ 
emission brightness ratio 
profiles, as a function of projected distance, presented in 
Figure~\ref{fig:emis-oxy-carb}. Since solar resonance fluorescence of 
atomic oxygen is an important excitation source for radial 
distances 
more than 20 km (see Figure~\ref{fig:emis-inten-ob}), the calculated 
O~{\scriptsize{I}} 1304/1356 \AA\ brightness ratio is 
increasing, whereas it is decreasing 
in case of O~{\scriptsize{I}} 1152/1304 \AA, with 
increasing projected distance. These 
calculations suggest that 
the observed O~{\scriptsize{I}} 1304/1356 \AA\ brightness 
ratio need not be 
an indicator to confirm the parent species of atomic oxygen 
emissions in the cometary coma.

We have modelled atomic oxygen 
density in the cometary coma by using hydrodynamical 
approach. But the global kinetic modelled number densities  {of 
\cite{Tenishev08}} close to the nucleus ($<$30 km) are higher about a factor 3
compared to our model calculated values.  {In order to evaluate  
the effect of solar resonant scattering of atomic oxygen, 
we have increased the atomic oxygen density in the model.}  By increasing
the atomic oxygen density about an order of magnitude, we 
find the calculated O~{\scriptsize{I}} 1304 \AA\ intensity 
increased by 50\% and oxygen emission ratio is in agreement 
with the observation when comet was at 2 AU. This 
calculation show that resonant scattering of atomic oxygen 
can also determine the oxygen emission ratio when 
substantial atomic oxygen is present in the inner coma. 
\cite{Feldman18} also noticed in the limb observation 
around the perihelion made on 2015 Aug 19 that resonance 
fluorescence is important excitation source in producing OI 
1304 \AA\ emission. 
Global kinetic models such as \cite{Tenishev08}, which 
accounts for strong momentum exchange between hot oxygen 
atoms to surrounding water molecules, can determine 
accurate atomic oxygen densities close to the nucleus which
is not possible by our one-dimensional approach.
However, our model calculations show that molecular 
oxygen 
is an important source of  O~{\scriptsize{I}}  emissions  
compared to H$_2$O via both photons and electron 
impact dissociative excitation. So we suggest that the 
observed atomic oxygen emissions are suitable to  derive 
O$_2$ relative abundances in the cometary coma.

\subsection{Limitations of model calculations}
The model calculations depend on several photons and 
photoelectron cross sections which have not been
critically evaluated in the literature. Most of the 
electron impact excitation cross sections are measured at 
either at 100 or 200 eV whereas the measurement of 
photodissociative excitation cross sections are limited to 
a small band of wavelengths, which leads to lesser 
dissociative excitation rates in the model. 
We could 
not  incorporate photodissociation channels of CO$_2$\ and 
CO due to non-availability of cross sections while calculating 
O {\scriptsize{I}} 1152 \AA\ emission intensity. Several 
assumptions have been made (see 
Section~\ref{sec:model-inputs}) to incorporate the various 
dissociative excitation channels producing atomic emission 
lines in the model. For the analysis of  {Alice} 
observations these calculations emphasize the urgent need 
for measuring various cross sections producing cometary 
spectroscopic emissions. Since model calculations are one 
dimensional, the heterogeneous distribution of neutrals in the 
cometary coma could not be incorporated to study the observed 
nadir emission intensities on different days.

\section{Summary and Conclusions} \label{summary}
The  {Alice}-ROSETTA observed  {various} O~{\scriptsize{I}} and 
H~{\scriptsize{I}} emissions {, when comet 67P/C-G was at 
large 
helioncentric distances ($>$3 AU).}  {These measured emission 
intensities}  have been used to 
derive electron densities in the coma assuming that 
electron impact dissociative excitation is the 
major source mechanism  {\citep{Feldman15}.} In order to quantify the 
contributions of photon and electron impact excitation 
processes, we have developed a photochemical 
model to study various  H~{\scriptsize{I}}, 
O~{\scriptsize{I}} and C~{\scriptsize{I}} emissions 
by 
accounting for essential formation mechanisms in the 
cometary coma.  {The following are major conclusions drawn 
from the present work.} 

\begin{enumerate}
\item Our model calculations on comet {67P/C-G}, {when it was 
	at 1.5 and 1.99 AU and having gas production rate of about 10$^{27}$ 
	s$^{-1}$,} 
show that photodissociative excitation reactions 
of primary neutrals significantly produce H~{\scriptsize{I}}, 
O~{\scriptsize{I}} and C~{\scriptsize{I}} emissions
compared to electron impact excitation processes.

\item   {Based on the model input parameters, 
	we suggest that }the photodissociation of 
O$_2$ is a significant
source of O~{\scriptsize{I}} emissions compared 
to that of H$_2$O. 

\item  {If our assumed photodissociation cross sections are realistic, 
model calculations show that the
	photodissociation of O$_2$ and H$_2$O can also be important sources of 
	O~{\scriptsize{I}} 1356 \AA\ emission 
	in the cometary coma rather}
than the earlier assumed electron impact excitation of O-bearing 
species. Hence, the observation 
of this emission line in comets not necessarily represents 
the presence of suprathermal electron impact excitation 
reactions. 

\item  {The photodissociative excitation of CO is a more significant   
	source in producing C~{\scriptsize{I}} emission lines compared to that of 
	CO$_2$. Hence, the observed C~{\scriptsize{I}} emission intensities are 
	suitable to derived CO abundances in the coma.}  
	
\item  {The observed H~{\scriptsize{I}} (Lyman-$\alpha$/Lyman-$\beta$), 
      O~{\scriptsize{I}}(1304/1356 \& 1152/1304), and 
      C~{\scriptsize{I}}(1657/1561) emission ratio vary based on the limb 
      viewing geometry and also depends on the excitation processes present 
      along the line of sight in the coma.}	

\item  {Solar resonance fluorescence is found to be important 
	 excitation source of these atomic emissions, except for O~{\scriptsize{I}} 
	 1356 \AA, for radial distances larger than 100 km in the coma. Hence, 
	 it is essential to study the atomic oxygen, carbon, and hydrogen 
	 distribution and their variation in the cometary coma to explain the 
	 observed emission ratios.}

\item When comet  {67P/C-G} is at large heliocentric distances  {($>$2 AU)} or 
      with a highly rarefied gaseous environment around the nucleus, 
      suprathermal electrons can not lose their energy to the surrounding 
      cometary volatile species which leads to a large population hot electrons 
      in the coma. In this case the electron impact could be a significant 
      source of atomic emissions hence the observed emission intensities 
      can be used to derive electron density in the coma.  But when comet has a 
      neutral gas production rate 10$^{27}$ s$^{-1}$ or more, 
      we  suggest that the observed H~{\scriptsize{I}}, O~{\scriptsize{I}}, and 
      C~{\scriptsize{I}} emissions are suitable to derive H$_2$O, O$_2$\ and CO 
      abundances.

\item Our modelled suprathermal electron flux is in agreement with 
      RPC-IES observations when the comet was at small
      heliocentric distances ($<$2 AU).      
\end{enumerate}


\section*{Acknowledgments}
The authors thank the anonymous reviewers for their valuable comments 
and suggestions that has improved the paper significantly. 
SR is supported by Department of Science and Technology 
(DST) with Innovation in
Science Pursuit for Inspired Research (INSPIRE) faculty award [Grant: DST/INSPIRE/04/2016/002687],
and he would like to thank Physical Research Laboratory for 
facilitating conducive research environment.







\begin{thebibliography}{80}
	\providecommand{\natexlab}[1]{#1}
	\expandafter\ifx\csname urlstyle\endcsname\relax
	\providecommand{\doi}[1]{doi:\discretionary{}{}{}#1}\else
	\providecommand{\doi}{doi:\discretionary{}{}{}\begingroup
		\urlstyle{rm}\Url}\fi
	
	\bibitem[{\textit{{Ajello}}(1971{\natexlab{a}})}]{Ajello71-co2}
	{Ajello}, J.~M. (1971{\natexlab{a}}), {Emission cross sections of 
	CO$_2$ by
		electron impact in the interval 1260-4500 {\AA}, II}, 
		\textit{J. Chem.
		Phys.}, \textit{55}, 3169--3177, \doi{10.1063/1.1676564}.
	
	\bibitem[{\textit{{Ajello}}(1971{\natexlab{b}})}]{Ajello71-co}
	{Ajello}, J.~M. (1971{\natexlab{b}}), {Emission cross Sections of 
	CO by
		electron impact in the interval 1260-5000 {\AA}. I}, 
		\textit{J. Chem. Phys.},
	\textit{55}, 3158--3168, \doi{10.1063/1.1676563}.
	
	\bibitem[{\textit{{Balsiger} et~al.}(2007)\textit{{Balsiger}, 
	{Altwegg},
			{Bochsler}, {Eberhardt}, {Fischer}, {Graf}, {J{\"a}ckel}, 
			{Kopp}, {Langer},
			{Mildner}, {M{\"u}ller}, {Riesen}, {Rubin}, {Scherer}, 
			{Wurz},
			{W{\"u}thrich}, {Arijs}, {Delanoye}, {de Keyser}, 
			{Neefs}, {Nevejans},
			{R{\`e}me}, {Aoustin}, {Mazelle}, {M{\'e}dale}, 
			{Sauvaud}, {Berthelier},
			{Bertaux}, {Duvet}, {Illiano}, {Fuselier}, {Ghielmetti}, 
			{Magoncelli},
			{Shelley}, {Korth}, {Heerlein}, {Lauche}, {Livi}, 
			{Loose}, {Mall}, {Wilken},
			{Gliem}, {Fiethe}, {Gombosi}, {Block}, {Carignan}, 
			{Fisk}, {Waite}, {Young},
			and {Wollnik}}}]{Balsiger07}
	{Balsiger}, H., K.~{Altwegg}, P.~{Bochsler}, P.~{Eberhardt}, 
	J.~{Fischer},
	S.~{Graf}, A.~{J{\"a}ckel}, E.~{Kopp}, U.~{Langer}, M.~{Mildner},
	J.~{M{\"u}ller}, T.~{Riesen}, M.~{Rubin}, S.~{Scherer}, P.~{Wurz},
	S.~{W{\"u}thrich}, E.~{Arijs}, S.~{Delanoye}, J.~{de Keyser}, 
	E.~{Neefs},
	D.~{Nevejans}, H.~{R{\`e}me}, C.~{Aoustin}, C.~{Mazelle}, J.-L. 
	{M{\'e}dale},
	J.~A. {Sauvaud}, J.-J. {Berthelier}, J.-L. {Bertaux}, L.~{Duvet}, 
	J.-M.
	{Illiano}, S.~A. {Fuselier}, A.~G. {Ghielmetti}, T.~{Magoncelli}, 
	E.~G.
	{Shelley}, A.~{Korth}, K.~{Heerlein}, H.~{Lauche}, S.~{Livi}, 
	A.~{Loose},
	U.~{Mall}, B.~{Wilken}, F.~{Gliem}, B.~{Fiethe}, T.~I. {Gombosi}, 
	B.~{Block},
	G.~R. {Carignan}, L.~A. {Fisk}, J.~H. {Waite}, D.~T. {Young}, and
	H.~{Wollnik} (2007), {Rosina Rosetta Orbiter Spectrometer for Ion 
	and Neutral
		Analysis}, \textit{Space Science Rev.}, \textit{128}, 
		745--801,
	\doi{10.1007/s11214-006-8335-3}.
	
	\bibitem[{\textit{{Beenakker} et~al.}(1974)\textit{{Beenakker}, 
	{Heer}, {Krop},
			and {M{\"o}hlmann}}}]{Beenakker74}
	{Beenakker}, C.~I.~M., F.~J.~D. {Heer}, H.~B. {Krop}, and G.~R. 
	{M{\"o}hlmann}
	(1974), {Dissociative excitation of water by electron impact},
	\textit{Chemical Physics}, \textit{6}, 445--454,
	\doi{10.1016/0301-0104(74)85028-7}.
	
	\bibitem[{\textit{Bhardwaj}(1999)}]{Bhardwaj99a}
	Bhardwaj, A. (1999), {On the role of solar EUV, photoelectrons, 
	and auroral
		electrons in the chemistry of C($^1$D) and the production of 
		C I 1931 {\AA}
		in the inner cometary coma: A case for comet P/Halley}, 
		\textit{J. Geophys.
		Res.}, \textit{104}, 1929 -- 1942, \doi{10.1029/1998JE900004}.
	
	\bibitem[{\textit{Bhardwaj}(2003)}]{Bhardwaj03}
	Bhardwaj, A. (2003), {On the solar EUV deposition in the inner 
	coma of comets
		with large gas production rates.}, \textit{Geophys. Res. 
		Lett.},
	\textit{30}(24), 2244, \doi{10.1029/2003GL018495}.
	
	\bibitem[{\textit{{Bhardwaj} and {Haider}}(2002)}]{Bhardwaj02}
	{Bhardwaj}, A., and S.~A. {Haider} (2002), {Chemistry of O($^1$D) 
	atoms in the
		coma: implications for cometary missions}, \textit{Adv. Space 
		Res.},
	\textit{29}, 745--750, \doi{10.1016/S0273-1177(02)00006-6}.
	
	\bibitem[{\textit{Bhardwaj and Jain}(2009)}]{Bhardwaj09}
	Bhardwaj, A., and S.~K. Jain (2009), {Monte Carlo model of 
	electron energy
		degradation in a CO$_2$ atmosphere}, \textit{J. Geophys. 
		Res.},
	\textit{114}(A13), 11,309, \doi{10.1029/2009JA014298}.
	
	\bibitem[{\textit{Bhardwaj and Michael}(1999)}]{Bhardwaj99b}
	Bhardwaj, A., and M.~Michael (1999), {On the excitation of Io's 
	atmosphere by
		the photoelectrons: Application of the analytical yield 
		spectrum of SO$_2$},
	\textit{Geophys. Res. Lett.}, \textit{26}, 393 -- 396,
	\doi{10.1029/1998GL900320}.
	
	\bibitem[{\textit{{Bhardwaj} and {Raghuram}}(2011)}]{Bhardwaj11}
	{Bhardwaj}, A., and S.~{Raghuram} (2011), {Model for Cameron-band 
	emission in
		comets: A case for the EPOXI mission target comet 
		103P/Hartley 2},
	\textit{Mon. Not. R. Astron. Soc.}, \textit{412}, L25 -- L29,
	\doi{10.1111/j.1745-3933.2010.00998.x}.
	
	\bibitem[{\textit{{Bhardwaj} and {Raghuram}}(2012)}]{Bhardwaj12}
	{Bhardwaj}, A., and S.~{Raghuram} (2012), {A coupled 
	chemistry-emission model
		for atomic oxygen green and red-doublet emissions in the 
		comet C/1996 B2
		Hyakutake}, \textit{Astrophys. J.}, \textit{748}, 13,
	\doi{10.1088/0004-637X/748/1/13}.
	
	\bibitem[{\textit{Bhardwaj et~al.}(1990)\textit{Bhardwaj, Haider, 
	and
			Singhal}}]{Bhardwaj90}
	Bhardwaj, A., S.~A. Haider, and R.~P. Singhal (1990), {Auroral and
		photoelectron fluxes in cometary ionospheres}, 
		\textit{Icarus}, \textit{85},
	216 -- 228, \doi{10.1016/0019-1035(90)90112-M}.
	
	\bibitem[{\textit{Bhardwaj et~al.}(1995)\textit{Bhardwaj, Haider, 
	and
			Singhal}}]{Bhardwaj95}
	Bhardwaj, A., S.~A. Haider, and R.~P. Singhal (1995), 
	{Consequences of cometary
		aurora on the carbon chemistry at comet P/Halley.}, 
		\textit{Adv. Space Res.},
	\textit{16 (2)}, 31 -- 36.
	
	\bibitem[{\textit{Bhardwaj et~al.}(1996)\textit{Bhardwaj, Haider, 
	and
			Singhal}}]{Bhardwaj96}
	Bhardwaj, A., S.~A. Haider, and R.~P. Singhal (1996), {Production 
	and emissions
		of atomic carbon and oxygen in the inner coma of comet 
		1P/Halley: Role of
		electron impact}, \textit{Icarus}, \textit{120}, 412 -- 430,
	\doi{10.1006/icar.1996.0061}.
	
	\bibitem[{\textit{{Bieler} et~al.}(2015)\textit{{Bieler}, 
	{Altwegg},
			{Balsiger}, {Bar-Nun}, {Berthelier}, {Bochsler}, 
			{Briois}, {Calmonte},
			{Combi}, {de Keyser}, {van Dishoeck}, {Fiethe}, 
			{Fuselier}, {Gasc},
			{Gombosi}, {Hansen}, {H{\"a}ssig}, {J{\"a}ckel}, {Kopp}, 
			{Korth}, {Le Roy},
			{Mall}, {Maggiolo}, {Marty}, {Mousis}, {Owen}, 
			{R{\`e}me}, {Rubin},
			{S{\'e}mon}, {Tzou}, {Waite}, {Walsh}, and 
			{Wurz}}}]{Bieler15}
	{Bieler}, A., K.~{Altwegg}, H.~{Balsiger}, A.~{Bar-Nun}, J.-J. 
	{Berthelier},
	P.~{Bochsler}, C.~{Briois}, U.~{Calmonte}, M.~{Combi}, J.~{de 
	Keyser}, E.~F.
	{van Dishoeck}, B.~{Fiethe}, S.~A. {Fuselier}, S.~{Gasc}, T.~I. 
	{Gombosi},
	K.~C. {Hansen}, M.~{H{\"a}ssig}, A.~{J{\"a}ckel}, E.~{Kopp}, 
	A.~{Korth},
	L.~{Le Roy}, U.~{Mall}, R.~{Maggiolo}, B.~{Marty}, O.~{Mousis}, 
	T.~{Owen},
	H.~{R{\`e}me}, M.~{Rubin}, T.~{S{\'e}mon}, C.-Y. {Tzou}, J.~H. 
	{Waite},
	C.~{Walsh}, and P.~{Wurz} (2015), {Abundant molecular oxygen in 
	the coma of
		comet 67P/Churyumov-Gerasimenko}, \textit{Nauture}, 
		\textit{526}, 678--681,
	\doi{10.1038/nature15707}.
	
	\bibitem[{\textit{{Bodewits} et~al.}(2016)\textit{{Bodewits}, 
	{Lara},
			{A'Hearn}, {La Forgia}, {Gicquel}, {Kovacs}, 
			{Knollenberg}, {Lazzarin},
			{Lin}, {Shi}, {Snodgrass}, {Tubiana}, {Sierks}, 
			{Barbieri}, {Lamy},
			{Rodrigo}, {Koschny}, {Rickman}, {Keller}, {Barucci}, 
			{Bertaux}, {Bertini},
			{Boudreault}, {Cremonese}, {Da Deppo}, {Davidsson}, 
			{Debei}, {De Cecco},
			{Fornasier}, {Fulle}, {Groussin}, {Guti{\'e}rrez}, 
			{G{\"u}ttler}, {Hviid},
			{Ip}, {Jorda}, {Kramm}, {K{\"u}hrt}, {K{\"u}ppers}, 
			{L{\'o}pez-Moreno},
			{Marzari}, {Naletto}, {Oklay}, {Thomas}, {Toth}, and 
			{Vincent}}}]{Bodewits16}
	{Bodewits}, D., L.~M. {Lara}, M.~F. {A'Hearn}, F.~{La Forgia}, 
	A.~{Gicquel},
	G.~{Kovacs}, J.~{Knollenberg}, M.~{Lazzarin}, Z.~Y. {Lin}, 
	X.~{Shi},
	C.~{Snodgrass}, C.~{Tubiana}, H.~{Sierks}, C.~{Barbieri}, P.~L. 
	{Lamy},
	R.~{Rodrigo}, D.~{Koschny}, H.~{Rickman}, H.~U. {Keller}, M.~A. 
	{Barucci},
	J.~L. {Bertaux}, I.~{Bertini}, S.~{Boudreault}, G.~{Cremonese}, 
	V.~{Da
		Deppo}, B.~{Davidsson}, S.~{Debei}, M.~{De Cecco}, 
		S.~{Fornasier},
	M.~{Fulle}, O.~{Groussin}, P.~J. {Guti{\'e}rrez}, 
	C.~{G{\"u}ttler}, S.~F.
	{Hviid}, W.~H. {Ip}, L.~{Jorda}, J.~R. {Kramm}, E.~{K{\"u}hrt},
	M.~{K{\"u}ppers}, J.~J. {L{\'o}pez-Moreno}, F.~{Marzari}, 
	G.~{Naletto},
	N.~{Oklay}, N.~{Thomas}, I.~{Toth}, and J.~B. {Vincent} (2016), 
	{Changes in
		the Physical Environment of the Inner Coma of 
		67P/Churyumov-Gerasimenko with
		Decreasing Heliocentric Distance}, \textit{Astronomical 
		Journal},
	\textit{152}(5), 130, \doi{10.3847/0004-6256/152/5/130}.
	
	\bibitem[{\textit{{Broiles} et~al.}(2016)\textit{{Broiles}, 
	{Burch}, {Chae},
			{Clark}, {Cravens}, {Eriksson}, {Fuselier}, {Frahm}, 
			{Gasc}, {Goldstein},
			{Henri}, {Koenders}, {Livadiotis}, {Mandt}, {Mokashi}, 
			{Nemeth}, {Odelstad},
			{Rubin}, and {Samara}}}]{Broiles16}
	{Broiles}, T.~W., J.~L. {Burch}, K.~{Chae}, G.~{Clark}, T.~E. 
	{Cravens},
	A.~{Eriksson}, S.~A. {Fuselier}, R.~A. {Frahm}, S.~{Gasc}, 
	R.~{Goldstein},
	P.~{Henri}, C.~{Koenders}, G.~{Livadiotis}, K.~E. {Mandt}, 
	P.~{Mokashi},
	Z.~{Nemeth}, E.~{Odelstad}, M.~{Rubin}, and M.~{Samara} (2016), 
	{Statistical
		analysis of suprathermal electron drivers at 
		67P/Churyumov-Gerasimenko},
	\textit{Mon. Not. R. Astron. Soc.}, \textit{462}, S312--S322,
	\doi{10.1093/mnras/stw2942}.
	
	\bibitem[{\textit{{Burch} et~al.}(2007)\textit{{Burch}, 
	{Goldstein}, {Cravens},
			{Gibson}, {Lundin}, {Pollock}, {Winningham}, and 
			{Young}}}]{Burch07}
	{Burch}, J.~L., R.~{Goldstein}, T.~E. {Cravens}, W.~C. {Gibson}, 
	R.~N.
	{Lundin}, C.~J. {Pollock}, J.~D. {Winningham}, and D.~T. {Young} 
	(2007),
	{RPC-IES: The Ion and Electron Sensor of the Rosetta Plasma 
	Consortium},
	\textit{Space Science Rev.}, \textit{128}(1-4), 697--712,
	\doi{10.1007/s11214-006-9002-4}.
	
	\bibitem[{\textit{Carlson}(1974)}]{Carlson74}
	Carlson, R.~W. (1974), Extreme ultraviolet photodissociative 
	excitation of
	molecular oxygen, \textit{The Journal of Chemical Physics}, 
	\textit{60}(6),
	2350--2353, \doi{10.1063/1.1681368}.
	
	\bibitem[{\textit{{Chaufray} et~al.}(2017)\textit{{Chaufray},
			{Bockel{\'e}e-Morvan}, {Bertaux}, {Erard}, {Feldman}, 
			{Capaccioni},
			{Schindhelm}, {Leyrat}, {Parker}, {Filacchione}, 
			{A'Hearn}, {Feaga},
			{Noonan}, {Keeney}, {Steffl}, {Stern}, {Weaver}, 
			{Broiles}, {Burch}, {Clark},
			and {Samara}}}]{Chaufray17}
	{Chaufray}, J.-Y., D.~{Bockel{\'e}e-Morvan}, J.-L. {Bertaux}, 
	S.~{Erard}, P.~D.
	{Feldman}, F.~{Capaccioni}, E.~{Schindhelm}, C.~{Leyrat}, 
	J.~{Parker},
	G.~{Filacchione}, M.~F. {A'Hearn}, L.~M. {Feaga}, J.~{Noonan}, 
	B.~{Keeney},
	A.~J. {Steffl}, S.~A. {Stern}, H.~A. {Weaver}, T.~{Broiles}, 
	J.~{Burch},
	G.~{Clark}, and M.~{Samara} (2017), {Rosetta Alice/VIRTIS 
	observations of the
		water vapour UV electroglow emissions around comet
		67P/Churyumov-Gerasimenko}, \textit{Monthly Notices of the 
		Royal Astronomical
		Society}, \textit{469}, S416--S426, 
		\doi{10.1093/mnras/stx1895}.
	
	\bibitem[{\textit{{Clark} et~al.}(2015)\textit{{Clark}, 
	{Broiles}, {Burch},
			{Collinson}, {Cravens}, {Frahm}, {Goldstein}, 
			{Goldstein}, {Mandt},
			{Mokashi}, {Samara}, and {Pollock}}}]{Clark15}
	{Clark}, G., T.~W. {Broiles}, J.~L. {Burch}, G.~A. {Collinson}, 
	T.~{Cravens},
	R.~A. {Frahm}, J.~{Goldstein}, R.~{Goldstein}, K.~{Mandt}, 
	P.~{Mokashi},
	M.~{Samara}, and C.~J. {Pollock} (2015), {Suprathermal electron 
	environment
		of comet 67P/Churyumov-Gerasimenko: Observations from the 
		Rosetta Ion and
		Electron Sensor}, \textit{Astron. \& Astrophys.}, 
		\textit{583}, A24,
	\doi{10.1051/0004-6361/201526351}.
	
	\bibitem[{\textit{Combi et~al.}(1998)\textit{Combi, Brown, 
	Feldman, Keller,
			Meier, and Smyth}}]{Combi98}
	Combi, M.~R., M.~E. Brown, P.~D. Feldman, H.~U. Keller, R.~R. 
	Meier, and W.~H.
	Smyth (1998), {Hubble space telescope ultraviolet imaging and 
	high-resolution
		spectroscopy of water photodissociation products in comet 
		Hyakutake (C/1996
		b2)}, \textit{Astrophys. J.}, \textit{494}, 816--821.
	
	\bibitem[{\textit{Combi et~al.}(2000)\textit{Combi, Reinard, 
	Bertaux,
			Quemerais, and M{\"a}kinen}}]{Combi00}
	Combi, M.~R., A.~A. Reinard, J.-L. Bertaux, E.~Quemerais, and 
	T.~M{\"a}kinen
	(2000), {SOHO/SWAN observations of the structure and evolution of 
	the
		Hydrogen Lyman-$\alpha$ coma of comet Hale-Bopp (1995 O1)}, 
		\textit{Icarus},
	\textit{144}, 191 -- 202, \doi{10.1006/icar.1999.6335}.
	
	\bibitem[{\textit{{Combi} et~al.}(2004)\textit{{Combi}, {Harris}, 
	and
			{Smyth}}}]{Combi04}
	{Combi}, M.~R., W.~M. {Harris}, and W.~H. {Smyth} (2004), 
	\textit{{Gas dynamics
			and kinetics in the cometary coma: theory and 
			observations}}, pp. 523--552,
	The university of arizon press.
	
	\bibitem[{\textit{{Edberg} et~al.}(2015)\textit{{Edberg}, 
	{Eriksson},
			{Odelstad}, {Henri}, {Lebreton}, {Gasc}, {Rubin}, 
			{Andr{\'e}}, {Gill},
			{Johansson}, {Johansson}, {Vigren}, {Wahlund}, {Carr}, 
			{Cupido},
			{Glassmeier}, {Goldstein}, {Koenders}, {Mandt}, {Nemeth}, 
			{Nilsson},
			{Richter}, {Wieser}, {Szego}, and {Volwerk}}}]{Edberg15}
	{Edberg}, N.~J.~T., A.~I. {Eriksson}, E.~{Odelstad}, P.~{Henri}, 
	J.~P.
	{Lebreton}, S.~{Gasc}, M.~{Rubin}, M.~{Andr{\'e}}, R.~{Gill}, 
	E.~P.~G.
	{Johansson}, F.~{Johansson}, E.~{Vigren}, J.~E. {Wahlund}, C.~M. 
	{Carr},
	E.~{Cupido}, K.~H. {Glassmeier}, R.~{Goldstein}, C.~{Koenders}, 
	K.~{Mandt},
	Z.~{Nemeth}, H.~{Nilsson}, I.~{Richter}, G.~S. {Wieser}, 
	K.~{Szego}, and
	M.~{Volwerk} (2015), {Spatial distribution of low-energy plasma 
	around comet
		67P/CG from Rosetta measurements}, \textit{Geophys. Res. 
		Lett.},
	\textit{42}(11), 4263--4269, \doi{10.1002/2015GL064233}.
	
	\bibitem[{\textit{Feldman et~al.}(2002)\textit{Feldman, Weaver, 
	and
			Burgh}}]{Feldman02}
	Feldman, P.~D., H.~A. Weaver, and E.~B. Burgh (2002), {Far 
	ultraviolet
		spectroscopic explorer observations of CO and H$_2$ emissions 
		in comet C/2001
		A2(Linear)}, \textit{Astrophys. J. Lett.}, \textit{576}, L91 
		-- L94.
	
	\bibitem[{\textit{Feldman et~al.}(2004)\textit{Feldman, Cochran, 
	and
			Combi}}]{Feldman04}
	Feldman, P.~D., A.~L. Cochran, and M.~R. Combi (2004), 
	\textit{{Spectroscopic
			investigations of fragment species in the coma: Comets 
			II}}, pp. 425--447, M.
	C. Festou, H. A. Weaver, \& H. U. Keller (Ed.)(Tucson: Univ. of 
	Arizona).
	
	\bibitem[{\textit{{Feldman} et~al.}(2015)\textit{{Feldman}, 
	{A'Hearn},
			{Bertaux}, {Feaga}, {Parker}, {Schindhelm}, {Steffl}, 
			{Stern}, {Weaver},
			{Sierks}, and {Vincent}}}]{Feldman15}
	{Feldman}, P.~D., M.~F. {A'Hearn}, J.-L. {Bertaux}, L.~M. 
	{Feaga}, J.~W.
	{Parker}, E.~{Schindhelm}, A.~J. {Steffl}, S.~A. {Stern}, H.~A. 
	{Weaver},
	H.~{Sierks}, and J.-B. {Vincent} (2015), {Measurements of the 
	near-nucleus
		coma of comet 67P/Churyumov-Gerasimenko with the Alice 
		far-ultraviolet
		spectrograph on Rosetta}, \textit{Astron. Astrophys.}, 
		\textit{583}, A8,
	\doi{10.1051/0004-6361/201525925}.
	
	\bibitem[{\textit{{Feldman} et~al.}(2016)\textit{{Feldman}, 
	{A'Hearn}, {Feaga},
			{Bertaux}, {Noonan}, {Parker}, {Schindhelm}, {Steffl}, 
			{Stern}, and
			{Weaver}}}]{Feldman16}
	{Feldman}, P.~D., M.~F. {A'Hearn}, L.~M. {Feaga}, J.-L. 
	{Bertaux}, J.~{Noonan},
	J.~W. {Parker}, E.~{Schindhelm}, A.~J. {Steffl}, S.~A. {Stern}, 
	and H.~A.
	{Weaver} (2016), {The Nature and Frequency of the Gas Outbursts 
	in Comet
		67P/Churyumov-Gerasimenko Observed by the Alice 
		Far-ultraviolet Spectrograph
		on Rosetta}, \textit{Astrophys. journal lett.}, \textit{825}, 
		L8,
	\doi{10.3847/2041-8205/825/1/L8}.
	
	\bibitem[{\textit{{Feldman} et~al.}(2018)\textit{{Feldman}, 
	{A'Hearn}, Bertaux,
			Feaga, Keeney, Knight, Noonan, Parker, Schindhelm, 
			Steffl, Stern, Vervack,
			and Weaver}}]{Feldman18}
	{Feldman}, P.~D., M.~F. {A'Hearn}, J.-L. Bertaux, L.~M. Feaga, 
	B.~A. Keeney,
	M.~M. Knight, J.~Noonan, J.~W. Parker, E.~Schindhelm, A.~J. 
	Steffl, S.~A.
	Stern, R.~J. Vervack, and H.~A. Weaver (2018), {FUV Spectral 
	Signatures of
		Molecules and the Evolution of the Gaseous Coma of Comet
		67P/Churyumov--Gerasimenko}, \textit{The Astronomical 
		Journal},
	\textit{155}(1), 9.
	
	\bibitem[{\textit{{Fougere} et~al.}(2016)\textit{{Fougere}, 
	{Altwegg},
			{Berthelier}, {Bieler}, {Bockel{\'e}e-Morvan}, 
			{Calmonte}, {Capaccioni},
			{Combi}, {De Keyser}, {Debout}, {Erard}, {Fiethe}, 
			{Filacchione}, {Fink},
			{Fuselier}, {Gombosi}, {Hansen}, {H{\"a}ssig}, {Huang}, 
			{Le Roy}, {Leyrat},
			{Migliorini}, {Piccioni}, {Rinaldi}, {Rubin}, {Shou}, 
			{Tenishev}, {Toth}, and
			{Tzou}}}]{Fougere16}
	{Fougere}, N., K.~{Altwegg}, J.-J. {Berthelier}, A.~{Bieler},
	D.~{Bockel{\'e}e-Morvan}, U.~{Calmonte}, F.~{Capaccioni}, M.~R. 
	{Combi},
	J.~{De Keyser}, V.~{Debout}, S.~{Erard}, B.~{Fiethe}, 
	G.~{Filacchione},
	U.~{Fink}, S.~A. {Fuselier}, T.~I. {Gombosi}, K.~C. {Hansen},
	M.~{H{\"a}ssig}, Z.~{Huang}, L.~{Le Roy}, C.~{Leyrat}, 
	A.~{Migliorini},
	G.~{Piccioni}, G.~{Rinaldi}, M.~{Rubin}, Y.~{Shou}, 
	V.~{Tenishev}, G.~{Toth},
	and C.-Y. {Tzou} (2016), {Direct Simulation Monte Carlo modelling 
	of the
		major species in the coma of comet 
		67P/Churyumov-Gerasimenko}, \textit{Mon.
		Not. R. Astron. Soc.}, \textit{462}, S156--S169, 
		\doi{10.1093/mnras/stw2388}.
	
	\bibitem[{\textit{{Galand} et~al.}(2016)\textit{{Galand}, 
	{H{\'e}ritier},
			{Odelstad}, {Henri}, {Broiles}, {Allen}, {Altwegg}, 
			{Beth}, {Burch}, {Carr},
			{Cupido}, {Eriksson}, {Glassmeier}, {Johansson}, 
			{Lebreton}, {Mandt},
			{Nilsson}, {Richter}, {Rubin}, {Sagni{\`e}res}, 
			{Schwartz}, {S{\'e}mon},
			{Tzou}, {Valli{\`e}res}, {Vigren}, and {Wurz}}}]{Galand16}
	{Galand}, M., K.~L. {H{\'e}ritier}, E.~{Odelstad}, P.~{Henri}, 
	T.~W. {Broiles},
	A.~J. {Allen}, K.~{Altwegg}, A.~{Beth}, J.~L. {Burch}, C.~M. 
	{Carr},
	E.~{Cupido}, A.~I. {Eriksson}, K.-H. {Glassmeier}, F.~L. 
	{Johansson}, J.-P.
	{Lebreton}, K.~E. {Mandt}, H.~{Nilsson}, I.~{Richter}, 
	M.~{Rubin}, L.~B.~M.
	{Sagni{\`e}res}, S.~J. {Schwartz}, T.~{S{\'e}mon}, C.-Y. {Tzou},
	X.~{Valli{\`e}res}, E.~{Vigren}, and P.~{Wurz} (2016), 
	{Ionospheric plasma of
		comet 67P probed by Rosetta at 3 au from the Sun}, 
		\textit{Mon. Not. R.
		Astron. Soc.}, \textit{462}, S331--S351, 
		\doi{10.1093/mnras/stw2891}.
	
	\bibitem[{\textit{{Gasc} et~al.}(2017)\textit{{Gasc}, {Altwegg}, 
	{Balsiger},
			{Berthelier}, {Bieler}, {Calmonte}, {Fiethe}, {Fuselier}, 
			{Galli}, and
			{Gombosi}}}]{Gasc17b}
	{Gasc}, S., K.~{Altwegg}, H.~{Balsiger}, J.-J. {Berthelier}, 
	A.~{Bieler},
	U.~{Calmonte}, B.~{Fiethe}, S.~{Fuselier}, A.~{Galli}, and 
	T.~{Gombosi}
	(2017), {Change of outgassing pattern of 
	67P/Churyumov-Gerasimenko during the
		March 2016 equinox as seen by ROSINA}, \textit{Monthly 
		Notices of the Royal
		Astronomical Society}, \textit{469}, S108--S117, 
		\doi{10.1093/mnras/stx1412}.
	
	\bibitem[{\textit{{Gladstone} et~al.}(2010)\textit{{Gladstone}, 
	{Hurley},
			{Retherford}, {Feldman}, {Pryor}, {Chaufray}, {Versteeg}, 
			{Greathouse},
			{Steffl}, {Throop}, {Parker}, {Kaufmann}, {Egan}, 
			{Davis}, {Slater},
			{Mukherjee}, {Miles}, {Hendrix}, {Colaprete}, and 
			{Stern}}}]{Gladstone10}
	{Gladstone}, G.~R., D.~M. {Hurley}, K.~D. {Retherford}, P.~D. 
	{Feldman}, W.~R.
	{Pryor}, J.-Y. {Chaufray}, M.~{Versteeg}, T.~K. {Greathouse}, 
	A.~J. {Steffl},
	H.~{Throop}, J.~W. {Parker}, D.~E. {Kaufmann}, A.~F. {Egan}, 
	M.~W. {Davis},
	D.~C. {Slater}, J.~{Mukherjee}, P.~F. {Miles}, A.~R. {Hendrix},
	A.~{Colaprete}, and S.~A. {Stern} (2010), {LRO-LAMP Observations 
	of the
		LCROSS Impact Plume}, \textit{Science}, \textit{330}, 472,
	\doi{10.1126/science.1186474}.
	
	\bibitem[{\textit{Haddad and Samson}(1986)}]{Haddad86}
	Haddad, G.~N., and J.~A.~R. Samson (1986), Total absorption and 
	photoionization
	cross sections of water vapor between 100 and 1000 \aa., 
	\textit{The Journal
		of Chemical Physics}, \textit{84}(12), 6623--6626, 
		\doi{10.1063/1.450715}.
	
	\bibitem[{\textit{Haider and Bhardwaj}(2005)}]{Haider05}
	Haider, S.~A., and A.~Bhardwaj (2005), {Radial distribution of 
	production
		rates, loss rates and densities corresponding to ion masses 
		$\le$40 amu in
		the inner coma of comet Halley: Composition and chemistry.}, 
		\textit{Icarus},
	\textit{177}, 196 -- 216, \doi{10.1016/j.icarus.2005.02.019}.
	
	\bibitem[{\textit{Hans et~al.}(2015)\textit{Hans, Knie, Schmidt, 
	Ben~Ltaief,
			Ozga, Rei\ss, Huckfeldt, F\"orstel, Hergenhahn, and 
			Ehresmann}}]{Hans15}
	Hans, A., A.~Knie, P.~Schmidt, L.~Ben~Ltaief, C.~Ozga, P.~Rei\ss, 
	H.~Huckfeldt,
	M.~F\"orstel, U.~Hergenhahn, and A.~Ehresmann (2015), 
	Lyman-series emission
	after valence and core excitation of water vapor, \textit{Phys. 
	Rev. A},
	\textit{92}, 032,511, \doi{10.1103/PhysRevA.92.032511}.
	
	\bibitem[{\textit{Hansen et~al.}(2016)\textit{Hansen, Altwegg, 
	Berthelier,
			Bieler, Biver, {Bockel{\'e}e-Morvan}, Calmonte, 
			Capaccioni, Combi, De~Keyser,
			Fiethe, Fougere, Fuselier, Gasc, Gombosi, Huang, Le~Roy, 
			Lee, Nilsson, Rubin,
			Shou, Snodgrass, Tenishev, Toth, Tzou, Simon~Wedlund, and 
			the
			ROSINA~team}}]{Hansen16}
	Hansen, K.~C., K.~Altwegg, J.-J. Berthelier, A.~Bieler, N.~Biver,
	D.~{Bockel{\'e}e-Morvan}, U.~Calmonte, F.~Capaccioni, M.~R. Combi,
	J.~De~Keyser, B.~Fiethe, N.~Fougere, S.~A. Fuselier, S.~Gasc, 
	T.~I. Gombosi,
	Z.~Huang, L.~Le~Roy, S.~Lee, H.~Nilsson, M.~Rubin, Y.~Shou, 
	C.~Snodgrass,
	V.~Tenishev, G.~Toth, C.-Y. Tzou, C.~Simon~Wedlund, and the 
	ROSINA~team
	(2016), {Evolution of water production of 
	67P/Churyumov--Gerasimenko: an
		empirical model and a multi-instrument study}, 
		\textit{Monthly Notices of the
		Royal Astronomical Society}, \textit{462}(Suppl), S491--S506,
	\doi{10.1093/mnras/stw2413}.
	
	\bibitem[{\textit{Haser}(1957)}]{Haser57}
	Haser, L. (1957), {Distribution d'intensite dans la tete d'une 
	comete},
	\textit{Bull. Acad. R Sci Liege}, \textit{43}, 740 -- 750.
	
	\bibitem[{\textit{{H{\"a}ssig} et~al.}(2015)\textit{{H{\"a}ssig}, 
	{Altwegg},
			{Balsiger}, {Bar-Nun}, {Berthelier}, {Bieler}, 
			{Bochsler}, {Briois},
			{Calmonte}, {Combi}, {De Keyser}, {Eberhardt}, {Fiethe}, 
			{Fuselier},
			{Galand}, {Gasc}, {Gombosi}, {Hansen}, {J{\"a}ckel}, 
			{Keller}, {Kopp},
			{Korth}, {K{\"u}hrt}, {Le Roy}, {Mall}, {Marty}, 
			{Mousis}, {Neefs}, {Owen},
			{R{\`e}me}, {Rubin}, {S{\'e}mon}, {Tornow}, {Tzou}, 
			{Waite}, and
			{Wurz}}}]{Hassig15}
	{H{\"a}ssig}, M., K.~{Altwegg}, H.~{Balsiger}, A.~{Bar-Nun}, J.~J.
	{Berthelier}, A.~{Bieler}, P.~{Bochsler}, C.~{Briois}, 
	U.~{Calmonte},
	M.~{Combi}, J.~{De Keyser}, P.~{Eberhardt}, B.~{Fiethe}, S.~A. 
	{Fuselier},
	M.~{Galand}, S.~{Gasc}, T.~I. {Gombosi}, K.~C. {Hansen}, 
	A.~{J{\"a}ckel},
	H.~U. {Keller}, E.~{Kopp}, A.~{Korth}, E.~{K{\"u}hrt}, L.~{Le 
	Roy},
	U.~{Mall}, B.~{Marty}, O.~{Mousis}, E.~{Neefs}, T.~{Owen}, 
	H.~{R{\`e}me},
	M.~{Rubin}, T.~{S{\'e}mon}, C.~{Tornow}, C.-Y. {Tzou}, J.~H. 
	{Waite}, and
	P.~{Wurz} (2015), {Time variability and heterogeneity in the coma 
	of
		67P/Churyumov-Gerasimenko}, \textit{Science}, 
		\textit{347}(1), aaa0276,
	\doi{10.1126/science.aaa0276}.
	
	\bibitem[{\textit{{Heritier} et~al.}(2018)\textit{{Heritier}, 
	{Altwegg},
			{Berthelier}, {Beth}, {Carr}, {Keyser}, {Eriksson}, 
			{Fuselier}, {Galand},
			{Gombosi}, {Henri}, {Nilsson}, {Rubin}~M., {Taylor}, and 
			E.}}]{Heritier18}
	{Heritier}, K., K.~{Altwegg}, J.-J. {Berthelier}, A.~{Beth}, 
	C.~{Carr}, J.~D.
	{Keyser}, A.~{Eriksson}, S.~{Fuselier}, M.~{Galand}, 
	T.~{Gombosi}, P.~J.~F.
	{Henri}, H.~{Nilsson}, S.~{Rubin}~M., {Wedlund}, M.~{Taylor}, and 
	V.~E.
	(2018), {On the origin of molecular oxygen in cometary comae}, 
	\textit{Nature
		Communications}, \textit{9}, 2580, 
		\doi{10.1038/s41467-018-04972-5}.
	
	\bibitem[{\textit{{Heritier} et~al.}(2017)\textit{{Heritier}, 
	{Altwegg},
			{Balsiger}, {Berthelier}, {Beth}, {Bieler}, {Biver}, 
			{Calmonte}, {Combi}, {De
				Keyser}, {Eriksson}, {Fiethe}, {Fougere}, {Fuselier}, 
				{Galand}, {Gasc},
			{Gombosi}, {Hansen}, {Hassig}, {Kopp}, {Odelstad}, 
			{Rubin}, {Tzou}, {Vigren},
			and {Vuitton}}}]{Heritier17}
	{Heritier}, K.~L., K.~{Altwegg}, H.~{Balsiger}, J.-J. 
	{Berthelier}, A.~{Beth},
	A.~{Bieler}, N.~{Biver}, U.~{Calmonte}, M.~R. {Combi}, J.~{De 
	Keyser}, A.~I.
	{Eriksson}, B.~{Fiethe}, N.~{Fougere}, S.~A. {Fuselier}, 
	M.~{Galand},
	S.~{Gasc}, T.~I. {Gombosi}, K.~C. {Hansen}, M.~{Hassig}, 
	E.~{Kopp},
	E.~{Odelstad}, M.~{Rubin}, C.-Y. {Tzou}, E.~{Vigren}, and 
	V.~{Vuitton}
	(2017), {Ion composition at comet 67P near perihelion: Rosetta 
	observations
		and model-based interpretation}, \textit{Mon. Not. R. Astron. 
		Soc.},
	\textit{469}, S427--S442, \doi{10.1093/mnras/stx1912}.
	
	\bibitem[{\textit{{Hoang} et~al.}(2017)\textit{{Hoang}, 
	{Altwegg}, {Balsiger},
			{Beth}, {Bieler}, {Calmonte}, {Combi}, {De Keyser}, 
			{Fiethe}, {Fougere},
			{Fuselier}, {Galli}, {Garnier}, {Gasc}, {Gombosi}, 
			{Hansen}, {J{\"a}ckel},
			{Korth}, {Lasue}, {Le Roy}, {Mall}, {R{\`e}me}, {Rubin}, 
			{S{\'e}mon},
			{Toublanc}, {Tzou}, {Waite}, and {Wurz}}}]{Hoang17}
	{Hoang}, M., K.~{Altwegg}, H.~{Balsiger}, A.~{Beth}, A.~{Bieler},
	U.~{Calmonte}, M.~R. {Combi}, J.~{De Keyser}, B.~{Fiethe}, 
	N.~{Fougere},
	S.~A. {Fuselier}, A.~{Galli}, P.~{Garnier}, S.~{Gasc}, 
	T.~{Gombosi}, K.~C.
	{Hansen}, A.~{J{\"a}ckel}, A.~{Korth}, J.~{Lasue}, L.~{Le Roy}, 
	U.~{Mall},
	H.~{R{\`e}me}, M.~{Rubin}, T.~{S{\'e}mon}, D.~{Toublanc}, C.-Y. 
	{Tzou}, J.~H.
	{Waite}, and P.~{Wurz} (2017), {The heterogeneous coma of comet
		67P/Churyumov-Gerasimenko as seen by ROSINA: H$_{2}$O, 
		CO$_{2}$, and CO from
		September 2014 to February 2016}, \textit{Astron. Astrophys}, 
		\textit{600},
	A77, \doi{10.1051/0004-6361/201629900}.
	
	\bibitem[{\textit{{Hoang} et~al.}(2019)\textit{{Hoang}, {Garnier},
			{Gourlaouen}, {Lasue}, {R{\`e}me}, {Altwegg}, {Balsiger}, 
			{Beth}, {Calmonte},
			{Fiethe}, {Galli}, {Gasc}, {J{\"a}ckel}, {Korth}, {Le 
			Roy}, {Mall}, {Rubin},
			{S{\'e}mon}, {Tzou}, {Waite}, and {Wurz}}}]{Hoang19}
	{Hoang}, M., P.~{Garnier}, H.~{Gourlaouen}, J.~{Lasue}, 
	H.~{R{\`e}me},
	K.~{Altwegg}, H.~{Balsiger}, A.~{Beth}, U.~{Calmonte}, 
	B.~{Fiethe},
	A.~{Galli}, S.~{Gasc}, A.~{J{\"a}ckel}, A.~{Korth}, L.~{Le Roy}, 
	U.~{Mall},
	M.~{Rubin}, T.~{S{\'e}mon}, C.~Y. {Tzou}, J.~H. {Waite}, and 
	P.~{Wurz}
	(2019), {Two years with comet 67P/Churyumov-Gerasimenko: 
	H$_{2}$O, CO$_{2}$,
		and CO as seen by the ROSINA/RTOF instrument of Rosetta}, 
		\textit{Astron. \&
		Astrophys.}, \textit{630}, A33, 
		\doi{10.1051/0004-6361/201834226}.
	
	\bibitem[{\textit{Huebner et~al.}(1992)\textit{Huebner, Keady, and
			Lyon}}]{Huebner92}
	Huebner, W.~F., J.~J. Keady, and S.~P. Lyon (1992), {Solar 
	photorates for
		planetary atmospheres and atmospheric pollutants}, 
		\textit{Astrophys. Space
		Sci.}, \textit{195}(1), 1--294, \doi{10.1007/BF00644558}.
	
	\bibitem[{\textit{Itikawa}(2002)}]{Itikawaco2}
	Itikawa, Y. (2002), {Cross sections for electron collisions with 
	carbon
		dioxide}, \textit{J. Phys. Chem. Ref. Data}, \textit{31}(3), 
		749--767,
	\doi{10.1063/1.1481879}.
	
	\bibitem[{\textit{Itikawa and Mason}(2005)}]{Itikawah2o}
	Itikawa, Y., and N.~Mason (2005), {Cross sections for electron 
	collisions with
		water molecules}, \textit{J. Phys. Chem. Ref. Data}, 
		\textit{34}(1), 1--22,
	\doi{10.1063/1.1799251}.
	
	\bibitem[{\textit{Kanik et~al.}(1993)\textit{Kanik, Ajello, and
			James}}]{Kanik93}
	Kanik, I., J.~M. Ajello, and G.~James (1993), Extreme ultraviolet 
	emission
	spectrum of co$_2$ induced by electron impact at 200 ev, 
	\textit{Chemical
		Physics Letters}, \textit{211}(6), 523 -- 528,
	\doi{10.1016/0009-2614(93)80137-E}.
	
	\bibitem[{\textit{Kanik et~al.}(1995)\textit{Kanik, James, and
			Ajello}}]{Kanik95}
	Kanik, I., G.~K. James, and J.~M. Ajello (1995), 
	Medium-resolution studies of
	extreme-ultraviolet emission from co by electron impact, 
	\textit{Phys. Rev.
		A}, \textit{51}, 2067--2074, \doi{10.1103/PhysRevA.51.2067}.
	
	\bibitem[{\textit{{Kanik} et~al.}(2003)\textit{{Kanik}, {Noren}, 
	{Makarov},
			{Vattipalle}, {Ajello}, and {Shemansky}}}]{Kanik03}
	{Kanik}, I., C.~{Noren}, O.~P. {Makarov}, P.~{Vattipalle}, J.~M. 
	{Ajello}, and
	D.~E. {Shemansky} (2003), {Electron impact dissociative 
	excitation of
		O$_{2}$: 2. Absolute emission cross sections of the OI(130.4 
		nm) and OI(135.6
		nm) lines}, \textit{J. Geophys. Res.}, \textit{108}, 5126,
	\doi{10.1029/2000JE001423}.
	
	\bibitem[{\textit{{Keeney} et~al.}(2017)\textit{{Keeney}, 
	{Stern}, {A'Hearn},
			{Bertaux}, {Feaga}, {Feldman}, {Medina}, {Parker}, 
			{Pineau}, {Schindhelm},
			{Steffl}, {Versteeg}, and {Weaver}}}]{Keeney17}
	{Keeney}, B.~A., S.~A. {Stern}, M.~F. {A'Hearn}, J.-L. {Bertaux}, 
	L.~M.
	{Feaga}, P.~D. {Feldman}, R.~A. {Medina}, J.~W. {Parker}, J.~P. 
	{Pineau},
	E.~{Schindhelm}, A.~J. {Steffl}, M.~{Versteeg}, and H.~A. 
	{Weaver} (2017),
	{H$_{2}$O and O$_{2}$ absorption in the coma of comet
		67P/Churyumov-Gerasimenko measured by the Alice 
		far-ultraviolet spectrograph
		on Rosetta}, \textit{Mon. Not. R. Astron. Soc.}, 
		\textit{469}, S158--S177,
	\doi{10.1093/mnras/stx1426}.
	
	\bibitem[{\textit{{Le Roy} et~al.}(2015)\textit{{Le Roy}, 
	{Altwegg},
			{Balsiger}, {Berthelier}, {Bieler}, {Briois}, {Calmonte}, 
			{Combi}, {De
				Keyser}, and {Dhooghe}}}]{Leroy15}
	{Le Roy}, L., K.~{Altwegg}, H.~{Balsiger}, J.-J. {Berthelier}, 
	A.~{Bieler},
	C.~{Briois}, U.~{Calmonte}, M.~R. {Combi}, J.~{De Keyser}, and 
	F.~{Dhooghe}
	(2015), {Inventory of the volatiles on comet 
	67P/Churyumov-Gerasimenko from
		Rosetta/ROSINA}, \textit{Astron. \& Astrophys.}, 
		\textit{583}, A1,
	\doi{10.1051/0004-6361/201526450}.
	
	\bibitem[{\textit{Lee et~al.}(1975)\textit{Lee, Carlson, Judge, 
	and
			Ogawa}}]{Lee75}
	Lee, L.~C., R.~W. Carlson, D.~L. Judge, and M.~Ogawa (1975), 
	Vacuum ultraviolet
	fluorescence from photodissociation fragments of co and co$_2$, 
	\textit{The
		Journal of Chemical Physics}, \textit{63}(9), 3987--3995,
	\doi{10.1063/1.431837}.
	
	\bibitem[{\textit{{Liu} and {Victor}}(1994)}]{Liu94}
	{Liu}, W., and G.~A. {Victor} (1994), {Electron energy deposition 
	in carbon
		monoxide gas}, \textit{Astrophys Journal}, \textit{435}, 
		909--919,
	\doi{10.1086/174872}.
	
	\bibitem[{\textit{{Madanian} et~al.}(2016)\textit{{Madanian}, 
	{Cravens},
			{Rahmati}, {Goldstein}, {Burch}, {Eriksson}, {Edberg}, 
			{Henri}, {Mandt},
			{Clark}, {Rubin}, {Broiles}, and {Reedy}}}]{Madanian16}
	{Madanian}, H., T.~E. {Cravens}, A.~{Rahmati}, R.~{Goldstein}, 
	J.~{Burch},
	A.~I. {Eriksson}, N.~J.~T. {Edberg}, P.~{Henri}, K.~{Mandt}, 
	G.~{Clark},
	M.~{Rubin}, T.~{Broiles}, and N.~L. {Reedy} (2016), {Suprathermal 
	electrons
		near the nucleus of comet 67P/Churyumov-Gerasimenko at 3 AU: 
		Model
		comparisons with Rosetta data}, \textit{Journal of 
		Geophysical Research
		(Space Physics)}, \textit{121}, 5815--5836, 
		\doi{10.1002/2016JA022610}.
	
	\bibitem[{\textit{{Madanian} et~al.}(2017)\textit{{Madanian}, 
	{Cravens},
			{Burch}, {Goldstein}, {Rubin}, {Nemeth}, {Goetz}, 
			{Koenders}, and
			{Altwegg}}}]{Madanian17}
	{Madanian}, H., T.~E. {Cravens}, J.~{Burch}, R.~{Goldstein}, 
	M.~{Rubin},
	Z.~{Nemeth}, C.~{Goetz}, C.~{Koenders}, and K.~{Altwegg} (2017), 
	{Plasma
		Environment around Comet 67P/Churyumov-Gerasimenko at 
		Perihelion: Model
		Comparison with Rosetta Data}, \textit{Astron. J.}, 
		\textit{153}, 30,
	\doi{10.3847/1538-3881/153/1/30}.
	
	\bibitem[{\textit{{Makarov} et~al.}(2003)\textit{{Makarov}, 
	{Kanik}, and
			{Ajello}}}]{Makarov03}
	{Makarov}, O.~P., I.~{Kanik}, and J.~M. {Ajello} (2003), 
	{Electron impact
		dissociative excitation of O$_{2}$: 1. Kinetic energy 
		distributions of fast
		oxygen atoms}, \textit{Journal of Geophysical Research 
		(Planets)},
	\textit{108}, 5125, \doi{10.1029/2000JE001422}.
	
	\bibitem[{\textit{{Makarov} et~al.}(2004)\textit{{Makarov}, 
	{Ajello},
			{Vattipalle}, {Kanik}, {Festou}, and 
			{Bhardwaj}}}]{Makarov04}
	{Makarov}, O.~P., J.~M. {Ajello}, P.~{Vattipalle}, I.~{Kanik}, 
	M.~C. {Festou},
	and A.~{Bhardwaj} (2004), {Kinetic energy distributions and line 
	profile
		measurements of dissociation products of water upon electron 
		impact},
	\textit{Journal of Geophysical Research (Space Physics)}, 
	\textit{109},
	A09303, \doi{10.1029/2002JA009353}.
	
	\bibitem[{\textit{{Meier}}(1995)}]{Meier95}
	{Meier}, R.~R. (1995), Solar lyman series line profiles and 
	atomic hydrogen
	excitation rates, \textit{Astrophys Journal}, \textit{452}, 462,
	\doi{10.1086/176318}.
	
	\bibitem[{\textit{{M{\"o}hlmann} 
	et~al.}(1978)\textit{{M{\"o}hlmann}, {Shima},
			and {De Heer}}}]{Mohlmann78}
	{M{\"o}hlmann}, G.~R., K.~H. {Shima}, and F.~J. {De Heer} (1978), 
	{Production
		of H, D(2s, 2p) by electron impact (0--2000 eV) on simple 
		hydrogen containing
		molecules}, \textit{Chemical Physics}, \textit{28}(3), 331 -- 
		341,
	\doi{10.1016/0301-0104(78)80010-X}.
	
	\bibitem[{\textit{{Morgan} and {Mentall}}(1974)}]{Morgan74}
	{Morgan}, H.~D., and J.~E. {Mentall} (1974), {VUV dissociative 
	excitation cross
		sections of H$_{2}$O, NH$_{3}$, and CH$_{4}$ by electron 
		impact}, \textit{J.
		Chem. phys.}, \textit{60}, 4734--4739, 
		\doi{10.1063/1.1680975}.
	
	\bibitem[{\textit{{Noonan} et~al.}(2018)\textit{{Noonan}, 
	{Stern}, {Feldman},
			{Broiles}, {Wedlund}, {Edberg}, {Schindhelm}, {Parker}, 
			{Keeney}, {Vervack},
			{Steffl}, {Knight}, {Weaver}, {Feaga}, {A'Hearn}, and 
			{Bertaux}}}]{Noonan18}
	{Noonan}, J.~W., S.~A. {Stern}, P.~D. {Feldman}, T.~{Broiles}, 
	C.~S. {Wedlund},
	N.~J.~T. {Edberg}, E.~{Schindhelm}, J.~W. {Parker}, B.~A. 
	{Keeney}, R.~J.
	{Vervack}, Jr., A.~J. {Steffl}, M.~M. {Knight}, H.~A. {Weaver}, 
	L.~M.
	{Feaga}, M.~{A'Hearn}, and J.-L. {Bertaux} (2018), {Ultraviolet 
	Observations
		of Coronal Mass Ejection Impact on Comet 
		67P/Churyumov-Gerasimenko by Rosetta
		Alice}, \textit{The Astronomical Journal}, \textit{156}, 16,
	\doi{10.3847/1538-3881/aac432}.
	
	\bibitem[{\textit{{Paxton}}(1985)}]{Paxton85}
	{Paxton}, L.~J. (1985), Pioneer venus orbiter ultraviolet 
	spectrometer limb
	observations - analysis and interpretation of the 166- and 156-nm 
	data,
	\textit{Journal of Geophys. Res.}, \textit{90}, 5089--5096,
	\doi{10.1029/JA090iA06p05089}.
	
	\bibitem[{\textit{Raghuram and Bhardwaj}(2013)}]{Raghuram13}
	Raghuram, S., and A.~Bhardwaj (2013), {Model for atomic oxygen 
	visible line
		emissions in comet C/1995 O1 Hale-Bopp}, \textit{Icarus}, 
		\textit{223},
	91--104, \doi{10.1016/j.icarus.2012.11.032}.
	
	\bibitem[{\textit{{Raghuram} et~al.}(2016)\textit{{Raghuram}, 
	{Bhardwaj}, and
			{Galand}}}]{Raghuram16}
	{Raghuram}, S., A.~{Bhardwaj}, and M.~{Galand} (2016), 
	{Prediction of Forbidden
		Ultraviolet and Visible Emissions in Comet 
		67P/Churyumov-Gerasimenko},
	\textit{Astrophys. J.}, \textit{818}, 102, 
	\doi{10.3847/0004-637X/818/2/102}.
	
	\bibitem[{\textit{Shirai et~al.}(2001)\textit{Shirai, Tabata, and
			Tawara}}]{Shirai01}
	Shirai, T., T.~Tabata, and H.~Tawara (2001), Analytic cross 
	sections for
	electron collisions with {CO}, {CO$_2$}, and {H$_2$O} relevant to 
	edge plasma
	impurities, \textit{Atomic Data and Nuclear Data Tables}, 
	\textit{79}(1), 143
	-- 184, \doi{10.1006/adnd.2001.0866}.
	
	\bibitem[{\textit{Singhal and Bhardwaj}(1991)}]{Singhal91}
	Singhal, R.~P., and A.~Bhardwaj (1991), {Monte Carlo Simulation of
		Photoelectron Energization in Parallel Electric Fields: 
		Electroglow on
		Uranus}, \textit{J. Geophys. Res.}, \textit{96}, 15,963 -- 
		15,972,
	\doi{10.1029/90JA02749}.
	
	\bibitem[{\textit{{Singhal} and {Haider}}(1984)}]{Singhal84}
	{Singhal}, R.~P., and S.~A. {Haider} (1984), {Analytical yield 
	spectrum
		approach to photoelectron fluxes in the earth's atmosphere}, 
		\textit{J.
		Geophys. Res.}, \textit{89}, 6847--6852, 
		\doi{10.1029/JA089iA08p06847}.
	
	\bibitem[{\textit{{Stern} et~al.}(2007)\textit{{Stern}, {Slater}, 
	{Scherrer},
			{Stone}, {Versteeg}, {A'Hearn}, {Bertaux}, {Feldman}, 
			{Festou}, {Parker}, and
			{Siegmund}}}]{Stern07}
	{Stern}, S.~A., D.~C. {Slater}, J.~{Scherrer}, J.~{Stone}, 
	M.~{Versteeg}, M.~F.
	{A'Hearn}, J.~L. {Bertaux}, P.~D. {Feldman}, M.~C. {Festou}, 
	J.~W. {Parker},
	and O.~H.~W. {Siegmund} (2007), {Alice: The rosetta Ultraviolet 
	Imaging
		Spectrograph}, \textit{Space Science Rev.}, \textit{128}, 
		507--527,
	\doi{10.1007/s11214-006-9035-8}.
	
	\bibitem[{\textit{{Tenishev} et~al.}(2008)\textit{{Tenishev}, 
	{Combi}, and
			{Davidsson}}}]{Tenishev08}
	{Tenishev}, V., M.~{Combi}, and B.~{Davidsson} (2008), {A Global 
	Kinetic Model
		for Cometary Comae: The Evolution of the Coma of the Rosetta 
		Target Comet
		Churyumov-Gerasimenko throughout the Mission}, 
		\textit{Astrophy. J.},
	\textit{685}, 659--677, \doi{10.1086/590376}.
	
	\bibitem[{\textit{Tozzi et~al.}(1998)\textit{Tozzi, Feldman, and
			Festou}}]{Tozzi98}
	Tozzi, G.~P., P.~D. Feldman, and M.~C. Festou (1998), {Origin and 
	production of
		C($^1$D) atoms in cometary comae}, \textit{Astron. 
		Astrophys.}, \textit{330},
	753--763.
	
	\bibitem[{\textit{Vigren and Galand}(2013)}]{Vigren13}
	Vigren, E., and M.~Galand (2013), Predictions of ion production 
	rates and ion
	number densities within the diamagnetic cavity of comet
	67p/churyumov-gerasimenko at perihelion, \textit{The 
	Astrophysical Journal},
	\textit{772}(1), 33.
	
	\bibitem[{\textit{{Woods} et~al.}(1986)\textit{{Woods}, 
	{Feldman}, {Dymond},
			and {Sahnow}}}]{Woods86}
	{Woods}, T.~N., P.~D. {Feldman}, K.~F. {Dymond}, and D.~J. 
	{Sahnow} (1986),
	{Rocket ultraviolet spectroscopy of comet Halley and abundance of 
	carbon
		monoxide and carbon}, \textit{Nature}, \textit{324}, 436--438,
	\doi{10.1038/324436a0}.
	
	\bibitem[{\textit{{Woods} et~al.}(2005)\textit{{Woods}, 
	{Eparvier}, {Bailey},
			{Chamberlin}, {Lean}, {Rottman}, {Solomon}, {Tobiska}, and
			{Woodraska}}}]{Woods05}
	{Woods}, T.~N., F.~G. {Eparvier}, S.~M. {Bailey}, P.~C. 
	{Chamberlin},
	J.~{Lean}, G.~J. {Rottman}, S.~C. {Solomon}, W.~K. {Tobiska}, and 
	D.~L.
	{Woodraska} (2005), {Solar EUV Experiment (SEE): Mission overview 
	and first
		results}, \textit{Journal of Geophysical Research (Space 
		Physics)},
	\textit{110}, A01312, \doi{10.1029/2004JA010765}.
	
	\bibitem[{\textit{{Wu} and {Judge}}(1979)}]{Wu79a}
	{Wu}, C.~Y.~R., and D.~L. {Judge} (1979), {The atomic oxygen 1304 
	{\AA}
		emission produced through photodissociation of CO and 
		CO$_{2}$.},
	\textit{Chemical Physics Letters}, \textit{68}, 495--498,
	\doi{10.1016/0009-2614(79)87245-0}.
	
	\bibitem[{\textit{Wu and Judge}(1981)}]{Wu81}
	Wu, C.~Y.~R., and D.~L. Judge (1981), Atomic carbon emission 
	produced through
	photodissociative excitation of {CO}, \textit{The Journal of 
	Chemical
		Physics}, \textit{75}(6), 2826--2830, \doi{10.1063/1.442355}.
	
	\bibitem[{\textit{{Wu} and {Judge}}(1988)}]{Wu88}
	{Wu}, C.~Y.~R., and D.~L. {Judge} (1988), Multichannel processes 
	of {H$_2$O} in
	the 18 {eV} region, \textit{The Journal of Chemical Physics},
	\textit{89}(10), 6275--6282, \doi{10.1063/1.455392}.
	
	\bibitem[{\textit{{Wu} et~al.}(1978)\textit{{Wu}, {Phillips}, 
	{Lee}, and
			{Judge}}}]{Wu78}
	{Wu}, C.~Y.~R., E.~{Phillips}, L.~C. {Lee}, and D.~L. {Judge} 
	(1978), {Atomic
		carbon emission from photodissociation of CO$_2$}, \textit{J. 
		Geophys. Res.},
	\textit{83}, 4869--4872.
	
	\bibitem[{\textit{Wu et~al.}(1979)\textit{Wu, Phillips, Lee, and 
	Judge}}]{Wu79}
	Wu, C. Y.~R., E.~Phillips, L.~C. Lee, and D.~L. Judge (1979), 
	Lyman-{$\alpha$}
	and balmer-series fluorescence from hydrogen photofragments of 
	{H$_2$O}
	vapor, \textit{The Journal of Chemical Physics}, \textit{70}(2), 
	601--608,
	\doi{10.1063/1.437539}.
	
	\bibitem[{\textit{Zhou et~al.}(2014)\textit{Zhou, Meng, and 
	Mo}}]{Zhou14}
	Zhou, Y., Q.~Meng, and Y.~Mo (2014), Photodissociation dynamics 
	of superexcited
	{O$_2$}: Dissociation channels {O}($^5$s) vs. {O}($^3$s), 
	\textit{The Journal
		of Chemical Physics}, \textit{141}(1), 014,301, 
		\doi{10.1063/1.4884906}.
	
\end{thebibliography}

\end{document}